\documentclass[11pt]{article}
\newcommand{\ShowComment}{true}
\newcommand{\floor}[1]{\lfloor #1 \rfloor}

\usepackage[utf8]{inputenc}

\usepackage{amsmath}
\usepackage{amsthm}
\usepackage{graphicx,epsfig}
\usepackage{caption}
\usepackage[noend]{algpseudocode}
\usepackage{titling}

\usepackage{amsmath}
\usepackage{amssymb}
\usepackage{amsthm}
\usepackage{amscd}
\usepackage{amsfonts}
\usepackage{graphicx}
\usepackage{fancyhdr}
\usepackage{bbold}
\usepackage{csquotes}

\usepackage{xcolor}
\usepackage[linesnumbered,ruled,vlined]{algorithm2e}
\usepackage{authblk}

\SetKwInput{KwInput}{Input}                % Set the Input
\SetKwInput{KwOutput}{Output}              % set the Output

\theoremstyle{plain} \numberwithin{equation}{section}
\newtheorem{theorem}{Theorem}[section]

\newtheorem{corollary}[theorem]{Corollary}

\newtheorem{lemma}[theorem]{Lemma}

\theoremstyle{plain}

\newtheorem{definition}[theorem]{Definition}

\newtheorem{claim}{Claim}[section]
\newcommand{\V}{\mathcal{V}}

\newcommand{\ceil}[1]{\left \lceil #1 \right \rceil}
\usepackage[backend=biber,style=numeric,sorting=nty,maxbibnames=99]{biblatex}

\advance \topmargin by -\headheight
\advance \topmargin by -\headsep
\evensidemargin \oddsidemargin
\ifdefined\ShowComment

	\newcommand{\thatchaphol}[1]{\textcolor{purple}{TS: #1}}

\else
	\newcommand{\seth}[1]{}
	\newcommand{\thatchaphol}[1]{}
	\newcommand{\longhui}[1]{}
\fi

\title{Deterministic Dynamic Matching in Worst-Case Update Time}

\author{Peter Kiss\footnote{This work is supported by Engineering and Physical Sciences Research Council, UK (EPSRC) Grant EP/S03353X/1}}

\affil{Department of Computer Science, University of Warwick, UK.}

\date{}

\begin{document}

\begin{titlingpage}
    \maketitle
    
\begin{abstract}

    We present deterministic algorithms for maintaining a $(3/2 + \epsilon)$ and $(2 + \epsilon)$-approximate maximum matching in a fully dynamic graph with worst-case update times $\hat{O}(\sqrt{n})$ and $\tilde{O}(1)$\footnote{Throughout the paper $\tilde{O}$ hides $poly( \log n, 1/\epsilon)$ factors and $\hat{O}$ hides $poly(n^{o(1)}, 1/\epsilon)$ factors.} respectively. The fastest known deterministic worst-case update time algorithms for achieving approximation ratio $(2 - \delta)$ (for any $\delta > 0$) and $(2 + \epsilon)$ were both shown by Roghani et al. [2021] with update times $O(n^{3/4})$ and $O_\epsilon(\sqrt{n})$ respectively. We close the gap between worst-case and amortized algorithms for the two approximation ratios as the best deterministic amortized update times for the problem are $O_\epsilon(\sqrt{n})$ and $\tilde{O}(1)$ which were shown in Bernstein and Stein [SODA'2021] and Bhattacharya and Kiss [ICALP'2021] respectively. 
    
    The algorithm achieving $(3/2 + \epsilon)$ approximation builds on the EDCS concept introduced by the influential paper of Bernstein and Stein [ICALP'2015]. Say that $H$ is a $(\alpha, \delta)$-approximate matching sparsifier if at all times $H$ satisfies that $\mu(H) \cdot \alpha  + \delta \cdot n \geq \mu(G)$ (define $(\alpha, \delta)$-approximation similarly for matchings). We show how to maintain a locally damaged version of the EDCS which is a $(3/2 + \epsilon, \delta)$-approximate matching sparsifier. We further show how to reduce the maintenance of an $\alpha$-approximate maximum matching to the maintenance of an $(\alpha, \delta)$-approximate maximum matching building based on an observation of Assadi et al. [EC'2016]. Our reduction requires an update time blow-up of $\hat{O}(1)$ or $\tilde{O}(1)$ and is deterministic or randomized against an adaptive adversary respectively. 
    
    To achieve $(2 + \epsilon)$-approximation we improve on the update time guarantee of an algorithm of Bhattacharya and Kiss [ICALP'2021]. In order to achieve both results we explicitly state a method implicitly used in Nanongkai and Saranurak [STOC'2017] and Bernstein et al. [arXiv'2020] which allows to transform dynamic algorithms capable of processing the input in batches to a dynamic algorithms with worst-case update time.

    \textbf{Independent Work:} Independently and concurrently to our work Grandoni et al. [arXiv'2021] has presented a fully dynamic algorithm for maintaining a $(3/2 + \epsilon)$-approximate maximum matching with deterministic worst-case update time $O_\epsilon(\sqrt{n})$.

\end{abstract}
\end{titlingpage}

\section{Introduction}

In the dynamic setting our task is to maintain a 'good' solution for some computational problem as the input undergoes updates \cite{m1,m3,m5,m13,m14,m17}. Our goal is to minimize the update time we need to spend in order to update the output when the input undergoes updates. One of the most extensively studied computational problems in the dynamic setting is approximate maximum matching. Our task is to maintain an $\alpha$-approximate matching $M$ in $G$, which is a matching which satisfies that $|M| \cdot \alpha \geq \mu(G)$ (where $\mu(G)$ represent the size of a maximum size matching of graph $G$). Due to the conditional lower bound of \cite{bound1} the maintenance of an exact maximum matching (a $1$-approximate maximum matching) requires at least $O(poly(n))$ update time. Hence, a long line of papers were focused on the possible approximation ratio-update time trade-offs achievable for $\alpha > 1$ \cite{m6,m7,m8,m9,m15,m16,bernstein2016faster,solomon2016fully}.

If a dynamic algorithm computes the updated output after at most $O(T)$ time following any single change in the input we say that its update time is worst-case $O(T)$. A slight relaxation of this bound is that the algorithm takes at most $O(T \cdot k)$ total time to maintain the output over $k>0$ consecutive updates to the input, for any k, in this case the update time of the algorithm is amortized $O(T)$.

A number of dynamic algorithms in literature utilize different levels of randomization \cite{sankowski2007faster, Arar18, Wajc20, m2, m4, m11}. However, currently all known techniques for proving update time lower bounds fail to differentiate between randomized and deterministic dynamic algorithms \cite{bound1, bound2, bound3, bound4, bound5}. Hence, understanding the power of randomization in the dynamic setting is an important research agenda. In the case of dynamic matching getting rid of randomization has proven to be difficult within the realm of $\tilde{O}(1)$ update time. While as early as the influential work of Onak and Rubinfield \cite{onak2010maintaining} a randomized algorithm with $\tilde{O}(1)$ update time has been found the first deterministic algorithm with the same update time was first shown by Bhattacharya et al. \cite{bhattacharya2016new}. For achieving $(2 + \epsilon)$-approximation with worst-case update time there is still an $O(poly(n))$ factor difference between the fastest randomized and deterministic implementations (\cite{Arar18, Wajc20} and \cite{roghani2021beating} respectively).

While amortized update time bounds don't tell us anything about worst-case update time some problems in the dynamic setting have proven to be difficult to solve efficiently without amortization. Notably, for the dynamic connectivity problem the first deterministic amortized update time solution by Holm et al. \cite{dynamicconnectivityfirstdeterministic} has long preceded the first worst-case update time implementation of Kapron et al. \cite{kapron2013dynamic} which required randomization.

Both of the algorithms presented by this paper carry the best of both worlds as they are deterministic and provide new worst-case update-time bounds. 

Many dynamic algorithms such as \cite{gupta2013fully, bhattacharya2021deterministic} rely on the robustness of the output of the output. To consider this in a context of matching as an example observe that if a matching $M$ is $\alpha$-approximate it remains $(\alpha \cdot (1 + O(\epsilon)))$-approximate even after $\epsilon \cdot |M|$ edge updates. Hence, if we are to rebuild $M$ after the updates we can amortize its reconstruction cost over $\epsilon \cdot |M|$ time steps. However, such an approach initially inherently results in amortized update time bound. In some cases with additional technical effort de-amortization was shown to be achievable for these algorithms \cite{gupta2013fully, bernstein2020fully}. A natural question to ask is weather an amortized update time bound is always avoidable for amortized rebuild based dynamic algorithms. 

To answer this question we explicitly present a versatile framework for improving the update time bounds of amortized rebuild based algorithms to worst-case while incurring only a $\tilde{O}(1)$ blowup in update time. Our framework was implicitly shown in Bernstein et al. \cite{bernstein2020fully} and Nanongkai and Saranurak \cite{nanongkai2017dynamic}. To demonstrate the framework we present two new results: 

\newpage

\begin{theorem}

\label{th:polylogworstcase}

There is a deterministic algorithm for maintaining a $(2 + \epsilon)$-approximate matching in a fully dynamic graph with worst-case update time $O_\epsilon(\log^7(n)) = \tilde{O}(1)$ (where $O_{\epsilon}$ hides $O(poly(1/\epsilon)$ factors).

\end{theorem}

For the approximation ratio of $(2 + \epsilon)$ the best known worst-case update time algorithm of $\tilde{O}(\sqrt{n})$ was show recently in \cite{roghani2021beating}. However, $\tilde{O}(1)$ amortized update time algorithms were previously shown by \cite{bhattacharya2016new}, \cite{bhattacharya2021deterministic}. We show that an $O(poly(n))$ blowup in update time is not necessary to improve these bounds to worst-case. 

\begin{theorem}

\label{th:dmgEDCS}

There is a fully dynamic algorithm for maintaining a $(3/2 + \epsilon)$-approximate maximum matching in worst-case deterministic $\hat{O}(\frac{m}{n \cdot \beta} + \beta)$ (for our choice of $\beta$) or $\hat{O}(\sqrt{n})$ update time (where $\hat{O}$ hides $O(poly(n^{o(1)}, 1/\epsilon))$ factors).

\end{theorem}

For achieving better than than 2-approximation the fastest known worst-case update time of $\tilde{O}(\sqrt{n}\sqrt[8]{m})$ was shown in \cite{roghani2021beating}. Similar to the case of $(2 + \epsilon)$-approximation there is an $\tilde{O}(poly(n))$ faster algorithm achieving the same approximation ratio shown in \cite{bernstein2016faster} using amortization. We again show that such a large blowup is not necessary in order to achieve worst-case update times. 

In order to derive the later result we first show an amortized rebuild based algorithm for maintaining the widely utilized \cite{bernstein2015fully, bernstein2016faster, grandoni2021maintaining, assadi2019coresets, unified, streaming, weighted, streamingstronger} matching sparsifier EDCS introduced by Bernstein and Stein \cite{bernstein2015fully}. At the core of amortized rebuild based algorithms there is a static algorithm for efficiently recomputing the underlying data-structure. As the EDCS matching sparsifier (as far as we are aware) doesn't admit a deterministic near-linear time static algorithm, we introduce a relaxed version of the EDCS we refer to as \enquote{damaged EDCS}. For constructing a damaged EDCS we show a deterministic $\tilde{O}(m)$ static algorithm. Say that matching sparsifier (or matching) $H$ is $(\alpha, \delta)$-approximate if $\mu(H) \cdot \alpha + n \cdot \delta \geq \mu(G)$. A damaged EDCS is a $(3/2 + \epsilon, \delta)$-approximate matching sparsifier as opposed to the EDCS which is $(3/2+\epsilon)$-approximate. To counter this we show new reductions from $(\alpha + \epsilon)$ to $(\alpha, \delta)$-approximate dynamic matching algorithms based on ideas of \cite{behnezhad2020stochastic}, \cite{assadi2019stochastic}. Previous such reductions relied on the \textit{oblivious adversary} assumption that the input sequence is independent from the choices of the algorithm and is fixed beforehand. Our reductions work against an \textit{adaptive adversary} whose decisions may depend on the decisions and random bits of the algorithm. The update time blowup required by the reductions is $\tilde{O}(1)$ or $\hat{O}(1)$ if the reduction step is randomized or deterministic respectively. These reductions and the static algorithm for constructing a damaged EDCS might be of independent research interest. Using the randomized reduction we receive the following corollary:

\begin{corollary}

The update time bound of Theorem~\ref{th:dmgEDCS} can be improved to $\tilde{O}(\frac{m}{n \cdot \beta} + \beta)$ (or $\tilde{O}(\sqrt{n})$) if we allow for randomization against an adaptive adversary (where $\tilde{O}$ hides $O(poly(\log (n), 1/\epsilon))$ factors).

\end{corollary}

\subsection{Techniques}

We base our approach for improving an amortized rebuild based algorithm to worst-case update time on an observation implicitly stated in Bernstein et al. \cite{bernstein2020fully} (Lemma 6.1). Take an arbitrary input sequence of changes $I$ for a dynamic problem and arbitrarily partition it into $k$ continuous sub-sequences $I_i : i \in [k]$. If a dynamic algorithm with update time $O(T)$ is such that (knowing the partitionings) it can process the input sequence and the total time of processing sub-sequence $I_i$ is $O(|I_i| \cdot T)$ then call it $k$ batch-dynamic. Note that the update time guarantee of a batch-dynamic algorithm is stronger then of an amortized update time algorithm but it is weaker than a worst-case update time bound.

Building on the framework of \cite{bernstein2020fully} we show that an $O(\log(n))$ batch-dynamic algorithm $Alg$ can be used to maintain $\tilde{O}(1)$ parallel output tapes with \textit{worst-case update time} such that at all times at least one output tape contains a valid output of $Alg$ while only incurring a blowup of $\tilde{O}(1)$ in update-time. If $Alg$ is an $\alpha$-approximate dynamic matching algorithm then each of the $O(\log(n))$ output tapes each contain a matching. Therefore, the union of the output tapes is an $\alpha$-approximate matching sparsifier with maximum degree $O(\log(n))$ on which we can run the algorithm of Gupta and Peng \cite{gupta2013fully} to maintain an $(\alpha + \epsilon)$-approximate matching. 

Therefore, in order to find new worst-case update time dynamic matching algorithms we only have to find batch-dynamic algorithms. We show a framework (building on \cite{bernstein2020fully}) for transforming amortized rebuild based dynamic algorithms to batch-dynamic algorithms. On a high level an amortized rebuild based algorithm allows for a slack of $\epsilon$ factor damage to its underlying data-structure before commencing a rebuild. To turn such an algorithm $k$ batch-dynamic during the progressing of the $i$-th batch we ensure a slack of $\frac{i \cdot \epsilon}{k}$ instead. This way once the algorithm finishes processing a batch it has $\frac{\epsilon}{k}$ factor of slack it is allowed to take before commencing a rebuild meaning that the next rebuild operation is expected to happen well into the proceeding batch.

With this general method and some technical effort we show a batch-dynamic version of the $(2 + \epsilon)$-approximate dynamic matching algorithm of \cite{bhattacharya2021deterministic} and prove Theorem~\ref{th:polylogworstcase}. 

In order to generate a batch-dynamic algorithm for maintaining a $(3/2 + \epsilon)$-approximate maximum matching more work is required as algorithms currently present in literature for this approximation ratio are not conveniently amortized rebuild based. We introduce a relaxed version of the matching sparsifier EDCS (initially appeared in \cite{bernstein2015fully}) called 'damaged EDCS'. We further show that a damaged EDCS can be found in $\tilde{O}(m)$ time. We show that a damaged EDCS is robust against $\tilde{O}(n \cdot \beta)$ edge updates and has maximum degree $\beta$ for our choice of $\beta$. This means we can maintain the damaged EDCS in $\tilde{O}(\frac{m}{n \cdot \beta})$ amortized update time with periodic rebuilds. We can then run the algorithm of \cite{gupta2013fully} to maintain a matching in the damaged EDCS in $\tilde{O}(\beta)$ update time. 

\subsection{Independent Work}

Independently from our work Grandoni et al. \cite{grandoni2021maintaining} presented a dynamic algorithm for maintaining a $(3/2 + \epsilon)$-approximate matching with deterministic worst-case update time $O_\epsilon(m^{1/4})$, where $O_\epsilon$ is hiding $O(poly(1/\epsilon))$ dependency.

\section{Notations and Preliminaries}

Throughout this paper, we let $G = (V,E)$ denote the input graph and $n$ will stand for $|V|$ and $m$ will stand for the maximum of $|E|$ as the graph undergoes edge updates. $\deg_E(v)$ will stand for the degree of vertex $v$ in edge set $E$ while $N_E(v)$ stand for the set of neighbouring vertices of $v$ in edge set $E$. We will sometimes refer to $\deg_E(u) + \deg_E(v)$ as the degree of edge $(u,v)$ in $E$. A matching $M$ of graph $G$ is a subset of vertex disjoint edges of $E$. $\mu(G)$ refers to the size of a maximum cardinality matching of $G$. A matching $M$ is an $\alpha$-approximate maximum matching if $\alpha \cdot |M| \geq \mu(G)$. Define a matching to be $(\alpha, \delta)$-approximate if $|M| \cdot \alpha + \delta \cdot n \geq \mu(G)$. 

In the maximum dynamic matching problem the task is to maintain a large matching while the graph undergoes edge updates. In this paper we will be focusing on the fully dynamic setting where the graph undergoes both edge insertions and deletions over time. An algorithm is said to be a dynamic $\alpha$ (or $(\alpha, \delta)$)-approximate maximum matching algorithm if it maintains an $\alpha$ (or $(\alpha, \delta)$)-approximate matching at all times. A sub-graph $H \subseteq E$ is said to be an $\alpha$ (or $(\alpha, \delta)$)-approximate matching sparsifier if it contains an $\alpha$ (or $(\alpha, \delta)$)-approximate matching. We will  regularly be referring to the following influential result from literature:

\begin{lemma}

\label{lm:guptapeng}

Gupta and Peng \cite{gupta2013fully}: There is a $(1 + \epsilon)$-approximate maximum matching algorithm for fully dynamic graph $G$ with deterministic worst-case update time $O(\Delta/\epsilon^2)$ given the maximum degree of $G$ is at most $\Delta$ at all times.

\end{lemma}

Throughout the paper the notations $\tilde{O}(), \hat{O}()$ and $O_\epsilon()$ will be hiding $O(poly(\log(n), \epsilon))$, \\ $O(poly(n^{o(1)}, \epsilon))$ and $O(poly(\frac{1}{\epsilon}))$ factors from running times respectively. 

The update time of a dynamic algorithm is worst-case $O(T)$ if it takes at most $O(T)$ time for it to update the output each time the input undergoes a change. An algorithm update time is said to be amortized $O(T)$ if there is some integer $k>0$ such that over $k$ consecutive changes to the input the algorithm takes $O(k \cdot T)$ time steps to maintain the output. The recourse of a dynamic algorithm measures the changes the algorithm makes to its output per change to the input. Similarly to update time recourse can be amortized and worst-case.

We call a dynamic algorithm $k$ batch-dynamic with update time $O(T)$ if for any partitioning of the input sequence $I$ into $k$ sub-sequences $I_i : i \in [k]$ during the processing of $I$ the algorithm can process input sub-sequence $I_i$ in $O(T \cdot |I_i|)$ total update time. Note that this implies that the worst-case update time during the progressing of $I_i$ is $O(T \cdot |I_i|)$. The definition is based on \cite{bernstein2020fully}. A $k$-batch dynamic algorithm provides slightly better update time bounds then an amortized update time algorithm as we can select $k$ sub-sequences to amortize the update time over.

We will furthermore be referring to the following recent result from Solomon and Solomon \cite{recoursebound}: 

\begin{lemma}

\label{lm:recoursebound}

Theorem 1.3 of Solomon and Solomon \cite{recoursebound} (slightly phrased differently and trivially generalized for $(\alpha, \delta)$-approximate matchings): Any fully dynamic $\alpha$ (or $(\alpha, \delta)$)-approximate maximum matching algorithm with update time $O(T)$ can be transformed into an $(\alpha + \epsilon)$ (or $(\alpha + \epsilon,\delta)$)-approximate maximum matching algorithm with $O(T + \frac{\alpha}{\epsilon})$ update time and worst-case recourse of $O(\frac{\alpha}{\epsilon})$ per update. The update time of the new algorithm is worst-case if so is the underlying matching algorithm.

\end{lemma}

\begin{definition}

\label{def:NA}

Random variables $X_1,...,X_n$ are said to be negatively associated if for any non-decreasing functions $g,f$ and disjoint subsets $I,J \subseteq [n]$ we have that:

$$\text{Cov}(g(X_i : i \in I), h(X_j : j \in J)) \leq 0$$

\end{definition}

We will make use of the following influential result bounding the probability of a sum of negatively associated random variables falling far from their expectation.

\begin{lemma}

\label{lm:chernoff:NA}

(Chernoff bound for negatively associated random variables \cite{chernoff}): Let $\bar{X} = \sum_{i \in [n]} X_i$ where $X_i : i \in [n]$ are negatively associated and $\forall i \in [n] : X_i \in [0,1]$. Then for all $\delta \in (0,1)$:

$$\Pr[\bar{X} \leq (1 - \delta) \cdot \mathbb{E}[\bar{X}]] \leq exp(-\frac{\mathbb{E}[\bar{X}] \cdot \delta^2}{2})$$

\end{lemma}

\section{Batch Dynamic To Worst Case Update Time}

\label{sec:Batch Dynamic To Worst Case Update Time}

\subsection{$k$ Batch Amortized Update Time Dynamic Algorithm}

\begin{lemma}

\label{lm:batchtoworstcase:matching}

Given an $\alpha$ approximate (or $(\alpha, \epsilon)$-approximate) dynamic matching algorithm $Alg$ is $O(\log(n))$ batch-dynamic with update time $O(T(n))$ and dynamic graph $G$ undergoing edge insertions and deletions. There is an algorithm $Alg'$ which maintains $O(\log(n))$ matchings of $G$ such that at all times during progressing an input sequence of arbitrarily large polynomial length one of the matchings is $\alpha$ approximate (or $(\alpha, \epsilon)$-approximate). The update time of $Alg'$ is worst-case $O(T(n) \cdot \log^3(n))$ and it is deterministic if $Alg$ is deterministic. 

\end{lemma}

As this lemma was implicitly stated in \cite{bernstein2020fully} and \cite{nanongkai2017dynamic} in a less general setting we defer the proof to Appendix~\ref{app:batchtoworst}.

\begin{corollary}

\label{cor:batchamortized:matching}

If there exists an $\alpha$ (or $(\alpha, \delta)$)-approximate dynamic matching algorithm (where $\alpha = O(1)$) $Alg$ which is $O(\log(n))$ batch-dynamic with update time $O(T(n))$ then there is an $(\alpha + \epsilon)$ (or $(\alpha + \epsilon, \delta)$)-approximate matching algorithm $Alg'$ with worst case update time $O(\frac{T(n) \cdot \log^3 (n)}{\epsilon^3})$. If $Alg$ is deterministic so is $Alg'$.

\end{corollary}

\begin{proof}

Maintain $O(\log(n))$ parallel matchings of $G$ using the algorithm from Lemma~\ref{lm:batchtoworstcase:matching} in $O(T(n) \cdot \log^3(n))$ worst case update time. Their union, say $H$, is a a graph with maximum degree $O(\log(n))$ and is an $\alpha$ (or ($\alpha, \delta$))-approximate matching sparsifier and is a union of the output of $O(\log(n))$ dynamic matching algorithms with worst-case update time $O(T \cdot \log^2(n))$. By Lemma~\ref{lm:recoursebound} (\cite{recoursebound}) these approximate matching algorithms can be transformed into $(\alpha + \epsilon/2)$ (or $(\alpha + \epsilon/2, \delta)$)-approximate matching algorithms with $O(T \cdot \log^2(n) + \frac{\alpha}{\epsilon})$ update time and $O(\frac{\alpha}{\epsilon})$ worst-case recourse. This bounds the total recourse of the sparsifier at $O(\frac{\log (n) \cdot \alpha}{\epsilon})$. Therefore, with slack parameter $\frac{\epsilon}{2 \cdot \alpha}$ we can run the algorithm of Lemma~\ref{lm:guptapeng} (\cite{gupta2013fully}) to maintain an $(\alpha + \epsilon)$ (or $(\alpha + \epsilon, \delta)$)-approximate matching in the sparsifier with worst-case update time $O(T \cdot \log^3(n) + \frac{\log (n) \cdot \alpha}{\epsilon} + \frac{\log^2(n) \cdot \alpha^2}{\epsilon^3}) = O(\frac{T \cdot \log^3(n)}{\epsilon^3})$.

\end{proof}

Observe that the framework outlined by Lemma~\ref{lm:batchtoworstcase:matching}
has not exploited any property of the underlying batch-dynamic algorithm other than the nature of it's running time. This allows for a more general formulation of Lemma~\ref{lm:batchtoworstcase:matching}.

\begin{corollary}

\label{cor:batchamortized}

If there is a $O(\log(n))$ batch-dynamic algorithm $Alg$ with deterministic (randomized) update time $O(T(n))$ and $poly(n)$ length input update sequence $I$ then there is an algorithm $Alg'$ such that

\begin{itemize}
    \item The update time of $Alg'$ is worst-case deterministic (randomized) $O(T(n) \cdot \log^3(n))$ 
    \item $Alg'$ maintains $\log(n)$ parallel outputs and after processing update sequence $I[0,\tau)$ one of $Alg'$-s maintained outputs is equivalent to the output of $Alg$ after processing $I[0,\tau)$ partitioned into at most $\log(n)$ batches
\end{itemize}

\end{corollary}

\section{Vertex Set Sparsification}

\label{sec:Vertex Set Sparsification}

An $(\alpha, \delta)$-approximate matching sparsifier satisfies that $\mu(H) \cdot \alpha + n \cdot \delta \geq \mu(G)$. Selecting $\delta = \frac{\epsilon \cdot \mu(H)}{n}$ results in a $(\alpha + \epsilon)$-approximate sparsifier. The algorithm we present in this paper has a polynomial dependence on $1/\delta$ therefore we can't select the required $\delta$ value to receive an $(\alpha + \epsilon)$-approximate sparsifier assuming $\mu(H)$ is significantly lower then $\mu(G)$. To get around this problem we sparsify the vertex set to a size of $\hat{O}(\mu(H))$ while ensuring that the sparsified graph contains a matching of size $(1 - O(\epsilon)) \cdot \mu(G)$.

Let $V^k$ be a partitioning of the vertices of $G = (V,E)$ into $k$ sets $v^i : i \in [k]$. Define the concatenation of $G$ based on $V^k$ to be graph $G_{V^k}$ on $k$ vertices corresponding to vertex subsets $v^i$ where there is an edge between vertices $v^i$ and $v^j$ if and only if there is $u \in v^i$ and $w \in v^j$ such that $(u,w) \in E$. Note that maintaining $V^k$ as $G$ undergoes edge changes can be done in constant time. Also note that given a matching $M_{V^k}$ of $G_{V^k}$ is maintained under edge changes to $G_{V^k}$ in constant update time per edge changes to $M_{V^k}$ we can maintain a matching of the same size in $G$.

\subsection{Vertex Sparsification Against An Oblivious Adversary}

Assume we are aware of $\mu(G)$ (note we can guess $\mu(G)$ within an $1+\epsilon$ multiplicative factor through running $O(\frac{\log(n)}{\epsilon})$ parallel copies of the algorithm). Choose a partitioning of $G$-s vertices into $O(\mu(G)/\epsilon)$ vertex subsets $V'$ uniformly at random. Define $G'$ to be the concatenation of $G$ based on $V'$.

\medskip

Consider a maximum matching $M^*$ of $G$. It's edges have $2 \cdot \mu(G)$ endpoints. Fix a specific endpoint $v$. With probability $(1 - \frac{2 \cdot \mu(G)}{\mu(G)/\epsilon})^{2 \cdot \mu(G) - 1} \sim (1 - o(\epsilon))$ it falls in a vertex set of $V'$ no other endpoint of $M^*$ does. Hence, in expectation $2 \cdot \mu(G) \cdot (1 - O(\epsilon))$ endpoints of $M^*$ fall into unique vertex subsets of $V'$ with respect to other endpoints. This also implies that $\mu(G) \cdot (1 - O(\epsilon))$ edges of $M^*$ will have both of their endpoints falling into unique vertex sets of $V'$, hence $\mu(G') \geq \mu(G) \cdot (1 - O(\epsilon))$. This observation motivates the following lemma which can be concluded from \cite{assadi2019stochastic}, \cite{behnezhad2020stochastic} and \cite{gupta2013fully}.

\begin{lemma}

\label{lm:vertexsparsification:oblivious}

Assume there is a dynamic algorithm $Alg$ which maintains an $(\alpha, \delta)$-approximate maximum matching where $\alpha = O(1)$ in graph $G = (V,E)$ with update time $O(T(n, \delta))$. Then there is a randomized dynamic algorithm $Alg'$ which maintains an $(\alpha+\epsilon)$-approximate maximum matching in update time time $O(T(n, \epsilon^2) \cdot \frac{\log^2(n)}{\epsilon^4})$. If the running time of $Alg$ is worst-case (amortized) so will be the running time of $Alg'$.

(Stated without proof as it concludes from \cite{assadi2019stochastic}, \cite{behnezhad2020stochastic} \cite{gupta2013fully})

\end{lemma}

\subsection{Vertex Set Sparsification Using $(k,\epsilon)$ Matching Preserving Partitionings}

A slight disadvantage of the method described above is that if the adversary is aware of our selection of $V'$ they might insert a maximum matching within the vertices of a single subset in $V'$ which would be completely lost after concatenation. In order to counter this we will do the following: we will choose some $L$ different partitionings of the vertices in such a way that for any matching $M$ of $G$ most of $M$-s vertices fall into unique subsets in at least one partitioning. 

\begin{definition}

\label{def:matchingpreserving}

Call a set of partitionings $\V$ of the vertices of graph $G = (V,E)$ into $d$ vertex subsets is $(k,\epsilon)$ matching preserving if for any matching of size $k$ in $G$ there is a partitioning $V^{d}_i$ in $\V$ such that if $G'$ is a concatenation of $G$ based on $V'$ then $G'$ satisfies that $\mu(G') \geq (1 - \epsilon) \cdot k$.

\end{definition}

We will show that using randomization we can generate a $(k,\epsilon)$ matching preserving set of partitionings of size $O(\frac{\log(n)}{\epsilon^2})$ into $O(k/\epsilon)$ vertex subsets in polynomial time. Furthermore, we will show how to find an $(k,\epsilon)$ matching preserving set of partitionings of size $O(n^{(1)})$ into $O(k \cdot n^{o(1)})$ vertex subsets deterministically in polynomial time.

\begin{lemma}

\label{lm:vertexsparsification:general}

Assume there exists a dynamic matching algorithm $Alg_M$ maintaining an $(\alpha, \delta)$-approximate matching in update time $O(T(n,\delta))$ for $\alpha = O(1)$ as well as an algorithm $Alg_S$ generating an $(k,\epsilon)$ matching preserving set of vertex partitionings into $O(k \cdot C)$ vertex subsets of size $L$. Then there exists an algorithm $Alg$ maintaining an $(\alpha + \epsilon)$-approximate matching with update time $O(T(n,\epsilon/C) \cdot \frac{L^2 \cdot \log^2(n)}{\epsilon^4})$. If both $Alg_S$ and $Alg_M$ are deterministic then so is $Alg$. If $Alg_M$ is randomized against an adaptive adversary then so is $Alg$. If the update time of $Alg_M$ is worst-case then so is of $Alg$. $Alg$ makes a single call to $Alg_S$.

\end{lemma}

The proof of the lemma is deferred to Appendix~\ref{app:sparsification}. The intuition is as follows: through running $O(\frac{\log (n)}{\epsilon})$ parallel copies of the algorithm guess $\mu(G)$ within a $1 + \epsilon$ factor. In the knowledge of $\mu(G)$ run $Alg_M$ on the $L$ concatenations of $G$ we generate with $Alg_S$. Each of these concatenated sub-graphs are of size $O(\mu(G) \cdot C)$ and have maximum matching size $(1 - O(\epsilon)) \cdot \mu(G)$. Therefore running $Alg_M$ with $\delta$ parameter $\Theta(\epsilon/C)$ yields an $(\alpha + O(\epsilon))$-approximate matching in on of these $L$ graphs. Using the algorithm of \cite{gupta2013fully} find an approximate maximum matching in the union of the $O_\epsilon(L \cdot \log (n))$ concatenated graphs. Note that with an application of Lemma~\ref{lm:recoursebound} (\cite{recoursebound}) the update time can be changed into $O(\frac{T(n,\epsilon/C) \cdot L \cdot \log(n)}{\epsilon} + \frac{L^2 \cdot \log^2(n)}{\epsilon^5})$ as shown in the appendix.

\subsection{Generating Matching Preserving Partitionings Through Random Sampling}

\begin{lemma}

\label{lm:partition:randomized}

There is a randomized algorithm succeeding with $1 - 1/poly(n)$ probability for generating a $(k,\epsilon)$ matching preserving set of partitionings of graph $G$ into $O(k/\epsilon)$ vertex subsets of size $O(\frac{\log(n)}{\epsilon^2})$ running in polynomial time.

\end{lemma}

We defer the proof to Appendix~\ref{app:sparsification}. Essentially, $O(\frac{\log(n)}{\epsilon^2})$ random chosen vertex partitionings into $O(k/\epsilon)$ vertex subsets are $(k,\epsilon)$ matching preserving. Note, that in unbounded time we can find an appropriate set of partitionings deterministically as we can iterate through all possible sets of partitionings and test each separately.

\subsection{Generating Matching Preserving Partitionings Using Expanders}

We will define expander graphs as follows. Such expanders are sometimes called unbalanced or lossless expanders in literature.

\begin{definition}

\label{def:expander}

Define a $(k, d, \epsilon)$-expander graph as a bipartite graph $G = ((L,R),E)$ such that $\forall v \in L : deg_E(v) = d$ and for any $S \subseteq L$ such that $|S| \leq k$ we have that $|N_E(S)| \geq (1 - \epsilon) \cdot d \cdot |S|$.

\end{definition}

Graph expanders are extensively researched and have found a number of different applications. We will now show how an expander graph can be used to be the bases of an $(k,\epsilon)$ preserving set of partitionings. 

\begin{lemma}

\label{lm:partitioningfromexpander}

Assume there exists an algorithm $Alg$ which outputs a $(k, d, \epsilon)$-expander $G_{exp} = ((L,R),E)$ in $O(T(k,d,\epsilon))$ time. There is an algorithm $Alg'$ which outputs a set of $(k,\epsilon)$ matching preserving vertex partitionings of a vertex set of size $|L|$ into $|R|$ subsets of size $d$ with running time $O(T(k, d, \epsilon))$. $Alg'$ is deterministic if $Alg$ is deterministic.

\end{lemma}

\begin{proof}

Take graph $G = (V,E)$ and bipartite $(2 \cdot k,d,\epsilon/2)$ expander graph $G_{Exp} = ((V,R),E')$ such that vertices of the left partition of $G_{Exp}$ correspond to vertices of $V$. For each $v \in V$ define an arbitrary ordering of it's neighbours in $R$ according to $E'$ and let $N_{E'}(v)_i$ be it's $i$-th neighbour according to this ordering ($i \in [d]$). For each $i \in [d]$ and $v \in R$ define $V_{i,v} \subseteq V$ to be the set of vertices in $V$ whose $i$-th neighbour is $v$ (or $V_{i,v} = \{v' \in V : N_{E'}(v')_i = v \}$). 

Define set of vertex partitionings $\V =\{ V_i^{|R|} : i \in [d]\}$ where $V_i^{|R|}$ contain vertex sets $V_{i,v} : v \in R$. Fix a matching $M$ in $G$ of size $k$ and call it's endpoints $V_{M}$. By the definition of the expander we have that $|N_{E'}(V_{M})| \geq (1 - \epsilon/2) \cdot d \cdot 2 k$. Hence by the pigeonhole principle we have that $|N_{E'}(V_{M^*})_i| \geq (1 - \epsilon/2) \cdot 2 k$ for some $i \in [d]$. Define $G'$  as the concatenation of $G$ based on $V_i^{|R|}$. By the definition of $V_i^{|R|}$ at least $(1 - \epsilon/2) \cdot 2 k$ endpoints of $M$ are concatenated into vertices of $G'$ containing exactly one vertex of $V_M$. Therefore, $(1 - \epsilon) \cdot k$ edges of $M$ will have both their endpoints concatenated into unique vertices of $G'$ within $M$. Hence, $\mu(G') \geq (1 - \epsilon) \cdot k$ and $\V$ is a $(k,\epsilon)$ matching preserving set of partitionings.

\end{proof}

\begin{lemma}

\label{lm:expander:deterministic}

(Theorem 7.3 of \cite{capalbo2002randomness} and Proposition 7 of \cite{berinde2008combining}): Given $n \geq k$ and $\epsilon > 0$. There exists a $(k,d,\epsilon)$-expander graph $G_{exp} = ((L,R),E)$ such that $|L| = n$, $|R| = \frac{k \cdot 2^{O(\log^3(\log(n)/\epsilon))}}{poly(\epsilon)} = \hat{O}(k)$, $d = 2^{O(\log^3(\log(n)/\epsilon))} = \hat{O}(1)$ which can be deterministically computed in $\hat{O}(n)$ time.

\end{lemma}

\subsection{Black-Box Implications}

The following statements are black-box statements which can be concluded based on this section.

\begin{corollary}

\label{cor:oblivious}

\cite{assadi2019stochastic},\cite{behnezhad2020stochastic}: If there is a dynamic algorithm for maintaining an $(\alpha, \delta)$-approximate maximum matching for dynamic graphs in update time $O(T(n, \delta))$ then there is a randomized algorithm (against oblivious adversaries) for maintaining an $(\alpha + \epsilon)$-approximate maximum matching with update time $O(T(n,\epsilon^2) \cdot \frac{\log^2(n)}{\epsilon^4})$.

\end{corollary}

\begin{corollary}

\label{cor:adaptive}

If there is a dynamic algorithm for maintaining an $(\alpha, \delta)$-approximate maximum matching for dynamic graphs in update time $O(T(n, \delta))$ then there is a randomized algorithm for maintaining an $(\alpha + \epsilon)$-approximate maximum matching with update time $O(T(n, \epsilon^2) \cdot \frac{\log^4(n)}{\epsilon^8})$ which works against adaptive adversaries given the underlying algorithm also does.

\end{corollary}

\begin{proof}

Follows from Lemma~\ref{lm:vertexsparsification:general} and Lemma~\ref{lm:partition:randomized}.

\end{proof}

\begin{corollary}

\label{cor:deterministic}

If there is a dynamic algorithm for maintaining an $(\alpha, \delta)$-approximate maximum matching for dynamic graphs in update time $O(T(n, \delta))$ then there is a deterministic algorithm for maintaining an $(\alpha + \epsilon)$-approximate maximum matching with update time $\hat{O}(T(n,\frac{poly(\epsilon)}{n^ {o(1)}}))$ which is deterministic given the underlying matching algorithm is also deterministic.

\end{corollary}

\begin{proof}

Follows from Lemma~\ref{lm:vertexsparsification:general},  Lemma~\ref{lm:expander:deterministic} and Lemma~\ref{lm:partitioningfromexpander}.

\end{proof}

\section{$(3/2 + \epsilon)$-Approximate Fully Dynamic Matching In $\hat{O}(\sqrt{n})$ Worst-Case Deterministic Update Time}

\label{sec:damagedEDCS}

\subsection{Algorithm Outline}

In this section we present an amortized rebuild based algorithm for maintaining a locally relaxed EDCS we refer to as 'damaged EDCS'. The following definition and key-property originates from \cite{bernstein2015fully} and \cite{unified}.

\begin{definition}

From Bernstein and Stein \cite{bernstein2015fully}:

\label{def:EDCS}

Given graph $G = (V,E)$, $H \subseteq E$ is a $(\beta, \lambda)$-EDCS of $G$ if it satisfies that:

\begin{itemize}
    \item $\forall e \in H : deg_H(e) \leq \beta$
    \item $\forall e \in E\setminus H : deg_H(e) \geq \beta \cdot (1 - \lambda)$
\end{itemize}

\end{definition}

\begin{lemma}

\label{lm:EDCS:approximation}

From Assadi and Stein \cite{unified}:

If $\epsilon < 1/2$, $\lambda \leq \frac{\epsilon}{32}$, $\beta \geq 8 \cdot \lambda^2 \cdot \log(1/\lambda)$ and $H$ is a $(\beta, \lambda)$-EDCS of $G$ then $\mu(G) \leq \mu(H) \cdot (\frac{3}{2} + \epsilon)$

\end{lemma}

The intuition behind the algorithm is as follows: take a $(\beta, \lambda)$-EDCS $H$. Relax it's parameter bounds slightly through observing that $H$ is also a $(\beta \cdot (1 + \lambda), 4 \lambda)$-EDCS. As $H$ is a $(\beta, \lambda)$-EDCS for every edge $e$ in it's local neighbourhood $\Theta(\beta \cdot \lambda)$ edge updates may occur in an arbitrary fashion before either of the two edge degree bounds of a $(\beta \cdot (1 + \lambda), 4 \lambda)$-EDCS is violated on $e$. 

Therefore, after $\tilde{\Theta}(n \cdot \beta)$ edge updates the properties of a $(\beta \cdot (1 + \lambda), 4 \lambda)$-EDCS should only be violated in the local neighbourhood of $O(\delta \cdot n)$ vertices for some small $\delta$ of our choice. At this point the EDCS is locally 'damaged' and it's approximation ratio as a matching sparsifier is reduced to $(3/2 + O(\epsilon), \delta)$. However, the reductions appearing in Section~\ref{sec:Vertex Set Sparsification} allows us to improve this approximation ratio to $(3/2 + O(\epsilon))$. At this point we commence a rebuild, the cost of which can be amortized over $\tilde{\Theta}(n \cdot \beta)$ edge updates.

We then proceed to turn this amortized rebuild based algorithm into a batch-dynamic algorithm which we improve to worst-case update time using Lemma~\ref{lm:batchtoworstcase:matching}.

\subsection{Definition and Properties of  $(\beta, \lambda, \delta)$-Damaged EDCS}

In order to base an amortized rebuild based dynamic algorithm on the EDCS matching sparsifier we need an efficient algorithm for constructing an EDCS. As far as we are aware there is no known deterministic algorithm for constructing an EDCS in in $\hat{O}(n)$ time. In order to get around this we introduce a locally relaxed version of EDCS.

\begin{definition}

\label{def:damagedEDCS}

For graph $G = (V,E)$ a $(\beta, \lambda, \delta)$-damaged EDCS is a subset of edges $H \subseteq E$ such that there is a subset of 'damaged' vertices $V_D \subseteq V$ and the following properties hold:

\begin{itemize}
    \item $|V_D| \leq \delta \cdot |V|$
    \item $\forall e \in H : deg_H(e) \leq \beta$
    \item All $e \in E \setminus H$ such that $e \cap V_D = \emptyset$ satisfies $deg_H(e) \geq \beta \cdot (1 - \lambda)$
\end{itemize}

\end{definition}

\begin{lemma}

\label{lm:damagedEDCS:approximation}

If $\epsilon < 1/2$, $\lambda \leq \frac{\epsilon}{32}$, $\beta  \geq 8 \lambda^{-2} \log (1 / \lambda)$ and $H$ is a $(\beta, \lambda,\delta)$-damaged EDCS of graph $G = (V,E)$ then $H$ is an $(3/2 + \epsilon, \delta)$-approximate matching sparsifier.

\end{lemma}

\begin{proof}

Define the following edge-set: $E' = \{e \in E: e \cap V_D = \emptyset\}$. Observe, that $H$ is a $(\beta,\lambda)$-EDCS of $E' \cup H$. Fix a maximum matching $M^*$ of $G$. At least $\mu(G) - |V_D| = \mu(G) - \delta \cdot |V|$ edges of $M^*$ appear in $E'$ as each vertex of $V_D$ can appear on at most one edge of $M^*$. Therefore, $\mu((V,E')) \geq \mu(G) - |V| \cdot \delta$. Now the lemma follows from Lemma~\ref{lm:EDCS:approximation}.

\end{proof}

\subsection{Constructing A Damaged EDCS in Near-Linear Time}

\begin{algorithm}[H]
	\SetKwInput{KwInput}{Input}                
	\SetKwInput{KwOutput}{Output}              
	\DontPrintSemicolon
	\SetKwBlock{Repeat}{repeat}{}
	\label{alg:static:approxEDCS}
	\caption{StaticDamagedEDCS}
	
	\KwInput{$G = (V,E), \beta, \lambda, \delta$}
	\KwOutput{$H_{fin} \subseteq E$:$(\beta, \lambda, \delta)$-damaged EDCS of $G$}
	$H = \emptyset$
	
	\Repeat
	{
	    $E'$ = $\emptyset$ \;
	    \For{$e \in E/H$}
	    {
	        \If{$deg_H(e) < \beta \cdot (1 - \lambda/2)$}
	        {
	            $H \leftarrow H \cup \{e\}$ \;
	            $E' \leftarrow E' \cup \{e\}$ \;
	        }
	    }
	    \If{$|E'| \leq \frac{\delta \cdot \lambda \cdot \beta \cdot n}{16}$}
	    {
	        $V_D \leftarrow \{v \in V: deg_{E'}(v) >\frac{\lambda \cdot \beta}{8} \}$ \;
	        $E_D \leftarrow \{e \in E': |e \cap V_D| > 0\}$ \;
	        $H_{fin} \leftarrow H \setminus E_D$ \;
	        Return $H_{fin}$ \;
	    }
	    \For{$e \in H$}
	    {
	        \If{$deg_H(e) > \beta \cdot (1 - \lambda/4)$}
	        {
	            $H \leftarrow H \setminus \{e\}$
	        }
	    }
	}
	
\end{algorithm}

\begin{lemma}

\label{lm:static:damagedEDCS:correctness}

Algorithm~\ref{alg:static:approxEDCS} returns $H_{fin}$ as a $(\beta, \lambda, \delta)$-damaged EDCS of $G$.

\end{lemma}

The potential function $\Phi$ used in proof of the following lemma is based on \cite{bernstein2016faster}.

\begin{lemma}

\label{lm:static:damagedEDCS:runningtime}

Algorithm~\ref{alg:static:approxEDCS} runs in deterministic $O(\frac{m}{\delta \cdot \lambda^2})$ time.

\end{lemma}

The proofs of the lemmas are deferred to Appendix~\ref{app:EDCS}. The intuition is the following: at the start of each iteration we add all edges of the graph to $H$ which have $deg_H(e) < \beta \cdot (1 - \lambda/2)$. If we fail to add at least $O(\lambda \cdot \delta \cdot \beta \cdot n)$ such edges we terminate with $H$ stripped of some edges. At the end of each iteration we remove all edges such that $deg_H(e) > \beta$. Consider what happens if we fail to add $\Omega(\lambda \cdot \delta \cdot \beta \cdot n)$ edges in an iteration. That means that only in the local neighbourhood of $\Theta(\delta \cdot n)$ 'damaged' vertices could we have added $\Omega(\beta \cdot \lambda)$ edges in the last iteration. We strip away the edges around damaged vertices to get $H$. The running time argument is based on a potential function $\Phi$ from \cite{bernstein2016faster}. Initially it is $0$ and has an upper bound of $O(n \cdot \beta^2)$ and grows by at least $\Omega(n \cdot \beta^2 \cdot \lambda^2 \cdot \delta)$ in each iteration bounding the number of iterations by $O(\frac{1}{\delta \cdot \lambda^2})$.

\subsection{Maintaining a Damaged EDCS in $\tilde{O}(\frac{m}{n \cdot \beta})$ Update Time With Amortized Rebuilds}

\begin{algorithm}[H]
	\SetKwInput{KwInput}{Input}                
	\SetKwInput{KwOutput}{Output}              
	\DontPrintSemicolon
	\SetKwBlock{Repeat}{Initially and after every $\alpha$ edge updates}{}
	\label{alg:dynamic:approxEDCS}
	\caption{DynamicDamagedEDCS}
	
	\KwInput{$G = (V,E), \beta, \lambda, \delta$}
	\KwOutput{$H \subseteq E$:$(\beta, \lambda, \delta)$ damaged-EDCS of $G$}
	
	$\alpha \leftarrow \frac{n \cdot \delta\cdot \lambda \cdot \beta}{64}$ \;
	
	\Repeat{
	
	    $H \leftarrow$ StaticDamagedEDCS($G, \frac{\beta}{1 + \lambda/4}, \lambda/4, \delta/2$)\;
	    $E_D \leftarrow \emptyset$ \;
	    $E_I \leftarrow \emptyset$ \;
	}
	
	\SetKwFunction{FInsertEdge}{InsertEdge}
	\SetKwFunction{FDeleteEdge}{DeleteEdge}
	
	\SetKwProg{Fn}{Function}{:}{}
	\Fn{\FInsertEdge{(u,v)}}{
	    $E_I \leftarrow E_I \cup \{ (u,v)\}$\;
		\If{$\max \{deg_{E_I}(u), deg_{E_I}(v)\} < \frac{\beta \cdot \lambda}{16} - 1$ and $deg_H((u,v)) \leq \beta - 2$  }
		{
		    $H \leftarrow H \cup \{(u,v)\}$ \;
		}
	}
	
	\SetKwProg{Fn}{Function}{:}{}
	\Fn{\FDeleteEdge{e}}{
	    $E_D \leftarrow E_D \cup \{ e\}$ \;
		$H \leftarrow H \setminus \{e\}$ \;
	}
	
\end{algorithm}

Note that $E_D$ is defined for the purposes of the analysis.

\begin{lemma}

\label{lm:dynamic:approxEDCS:correctness}

The sparsifier $H$ maintained by Algorithm~\ref{alg:dynamic:approxEDCS} is a $(\beta, \lambda, \delta)$-damaged EDCS of $G$ whenever the algorithm halts. 

\end{lemma}

\begin{lemma}

\label{lm:dynamic:damagedEDCS:amortizedbounds}

The amortized update time of Algorithm~\ref{alg:dynamic:approxEDCS} over a series of $\alpha$ updates is $O(\frac{m}{n \cdot \beta \cdot \lambda^3 \cdot \delta^2})$ and the sparsifier $H$ undergoes $O(\frac{1}{\lambda \cdot \delta})$ amortized recourse.

\end{lemma}

The lemmas are proven in the appendix. On a high level, a $(\beta, O(\lambda), O(\delta))$ damaged EDCS will gain $O(n \cdot \delta)$ damaged vertices in the span of $O(n \cdot \delta \cdot \lambda \cdot \beta)$ edge updates as for a vertex to be damaged there has to be $O(\beta \cdot \lambda)$ edge updates in it's local neighbourhood. At this point we can call a rebuild of the EDCS in $\tilde{O}(m)$ time to get an amortized update time of $\tilde{O}(\frac{m}{\beta \cdot n})$.

\subsection{$k$ Batch-Dynamic Algorithm For Maintaining An Approximate EDCS}

\begin{lemma}

\label{lm:kbatchEDCS}

Given fully dynamic graph $G$ with $n$ vertices and $m$ edges. There is a $k$ batch-dynamic dynamic algorithm which maintains a $(\beta, \lambda, \delta)$-damaged EDCS of this graph with deterministic update time $O(\frac{k \cdot m}{n \cdot \beta \cdot \delta^2 \cdot \lambda^3})$ and recourse $O(\frac{k}{\delta \cdot \lambda})$.

\end{lemma}

\begin{proof}

Define an alternative version of Algorithm~\ref{alg:dynamic:approxEDCS} where $\alpha$ is simply set to $\alpha_i = i \cdot \frac{\alpha}{k}$ during the processing of the $i$-th batch. Observe that in the proof of Lemma~\ref{lm:dynamic:approxEDCS:correctness} the only detail which depends on the choice of $\alpha$ is the size of $V_{E_D} \cup V_{E_I}$. At any point in this batch modified version of the algorithm $\alpha_i \leq \alpha$ therefore the correctness of the algorithm follows.

The running time of the algorithm will be affected by this change. As every edge update is processed in constant time by the algorithm the running time is dominated by calls to StaticDamagedEDCS. By definition for every batch at least $\alpha/k$ edge updates will occur between the start of the batch and the first rebuild (if there is one) yielding an amortized update time of at most $O(\frac{k \cdot m}{n \cdot \beta \cdot \delta^2 \cdot \lambda^3})$ over the first rebuild (due to Lemma~\ref{lm:dynamic:damagedEDCS:amortizedbounds}). After the first rebuild the algorithm simply proceeds to run with $\alpha$-parameter $\alpha_i$ therefore the amortized update time for the remainder of batch $i$ is $O(\frac{i \cdot m}{n \cdot \beta \cdot \delta^2 \cdot \lambda^3}) = O(\frac{k \cdot m}{n \cdot \beta \cdot \delta^2 \cdot \lambda^3})$.

\end{proof}

\begin{corollary}

\label{cor:kbatchmatching}

For fully dynamic graph $G$ there is a deterministic $k$ batch-dynamic algorithm for maintaining a $(3/2+\epsilon, \delta)$-approximate maximum matching with update time $\tilde{O}(\frac{k \cdot m}{n \cdot \beta} + k \cdot \beta)$. 

\end{corollary}

\begin{proof}

Set $\lambda = \frac{\epsilon}{128}$ and $\beta$  large enough to satisfy the requirements of Lemma~\ref{lm:damagedEDCS:approximation} such that the resulting sparsifier is $(3/2 + \epsilon/4, \delta)$-approximate. Use the algorithm of Lemma~\ref{lm:kbatchEDCS}. The resulting damaged-EDCS sparsifier will have maximum degree $O(\beta)$, undergo $\tilde{O}(k)$ recourse per update and will take $\tilde{O}(\frac{k \cdot m}{n \cdot \beta})$ time to maintain. By Lemma~\ref{lm:damagedEDCS:approximation} it will be a $(3/2 + \epsilon/4, \delta)$-approximate matching sparsifier. Hence, if we apply the algorithm of Lemma~\ref{lm:guptapeng} to maintain a $(1 + \epsilon/4)$-approximate maximum matching within the sparsifier we can maintain a $(3/2 + \epsilon, \delta)$ approximate matching in $\tilde{O}(\frac{m \cdot k}{n \cdot \beta} + \beta \cdot k)$ update time and recourse.

\end{proof}

\subsection{Proof of Theorem~\ref{th:dmgEDCS}}

\begin{proof}

Take the algorithm of Corollary~\ref{cor:kbatchmatching}. Set $k = \log(n)$ and apply Corollary~\ref{cor:batchamortized:matching} to receive a deterministic $(3/2 + \epsilon, \delta)$-approximate dynamic matching algorithm with worst-case update time $\tilde{O}(\frac{m}{n \cdot \beta} + \beta)$. Finally, transform this algorithm into a $(3/2 + \epsilon)$-approximate matching algorithm using either Corollary~\ref{cor:adaptive} or Corollary~\ref{cor:deterministic}.

\end{proof}

\section{$(2+\epsilon)$-Approximate Fully Dynamic Maximum Matching in $\tilde{O}(1)$ Worst-Case Update Time}

\label{sec:BK-batchdynamic}

In the appendix we present a deterministic worst-case $O(poly(\log(n), 1/\epsilon))$-update time $(2 + \epsilon)$-approximate fully dynamic matching algorithm. Currently, the only deterministic $O(poly(\log(n), 1/\epsilon))$-update time algorithms \cite{bhattacharya2016new}, \cite{bhattacharya2021deterministic} have amortized update time bounds, while the fastest wort-case algorithm runs in $\tilde{O}(\sqrt{n})$ update time from \cite{roghani2021beating}. We will first improve the running time bounds of the algorithm presented in \cite{bhattacharya2021deterministic} to $k$ batch-dynamic using the same technique as presented previously. \cite{bhattacharya2021deterministic} similarly bases the algorithm on amortized rebuilds which are triggered when $\epsilon$ factor of change occurs within the data-structure. In order to improve the update time to batch-dynamic we define $\epsilon_i = \frac{i \cdot \epsilon}{k}$ to be the slack parameter during the progressing of batch $i$. Firstly, this ensures that $\epsilon_i \leq \epsilon$ during any of the batches progressed guaranteeing the approximation ratio. Secondly, whenever a new batch begins the slack parameter increases by $\frac{\epsilon}{k}$ which insures that there will be enough time steps before next rebuild occurs to amortize the rebuild time over.

\begin{lemma}

\label{lm:BKbatch}

There is a deterministic $k$ batch amortized $O_\epsilon(k \cdot \log^4(n))$ update time $(2+\epsilon)$-approximate fully dynamic matching algorithm.

\end{lemma}

\begin{proof}

Deferred to Appendix~\ref{app:BK-kbatch} as most of the argument is taken from \cite{bhattacharya2021deterministic}.

\end{proof}

Theorem~\ref{th:polylogworstcase} follows from Lemma~\ref{lm:BKbatch} and Corollary~\ref{cor:batchamortized:matching}.

\section{Acknowledgements}

\label{sec:acknowledgments}

We would like to thank Sayan Bhattacharya and Thatchaphol Saranurak for helpful discussions. Further we would like to thank Thatchaphol Saranurak for suggesting to use lossless expanders to deterministically generate $\epsilon$-matching preserving partitionings.

\section{Open Questions}

\textbf{Worst-Case Update Time Improvement Through Batch-Dynamization}: We have shown two applications on how batch-dynamization can be used to improve amortized rebuild based algorithm update times to worst-case. As amortized rebuild is a popular method for dynamizing a data-structure not just in the context of matching it would be interesting to see if the batch-dynamization based framework has any more applications.

\textbf{$(\alpha, \delta)$-Approximate Dynamic Matching}: In current dynamic matching literature most algorithms focus on maintaining an $\alpha$-approximate matching or matching sparsifier both for the integral and fractional version of the problem. However, a more relaxed $(\alpha, \delta)$-approximate matching algorithm using the reductions presented in this paper (or \cite{assadi2019stochastic}, \cite{behnezhad2020stochastic}) allow for the general assumption that $\mu(G) = \Theta(n)$ at all times. This assumption has proven to be useful in other settings for the matching problem such as the stochastic setting (\cite{assadi2019stochastic}, \cite{behnezhad2020stochastic}) but largely seems to be unexplored in the dynamic setting.

\textbf{Damaged EDCS}: The EDCS matching sparsifier \cite{bernstein2015fully} has found use in a number of different settings for the matching problem \cite{bernstein2016faster} \cite{streaming} \cite{streamingstronger} \cite{roghani2021beating} \cite{unified} \cite{weighted} \cite{assadi2019coresets}. In contrast with the EDCS (as far as we are aware) a damaged EDCS admits a deterministic near-linear time static algorithm. This might lead to new results in related settings.

\printbibliography

\appendix

\section{Proof of Lemma~\ref{lm:batchtoworstcase:matching}}

\label{app:batchtoworst}

We restate the lemma for the readers convenience:

\begin{lemma}

Given $(\alpha)$ approximate (or $(\alpha, \epsilon)$-approximate) dynamic matching algorithm $Alg$ is $O(\log(n))$ batch-dynamic with update time $O(T(n))$ and dynamic graph $G$ undergoing edge insertions and deletions. There is an algorithm $Alg'$ which maintains $O(\log(n))$ matchings of $G$ such that at all times during processing an input sequence of arbitrarily large polynomial length one of the matchings is $(\alpha)$ approximate (or $(\alpha, \epsilon)$-approximate). The update time of $Alg'$ is worst-case $O(T(n) \cdot \log^3(n))$ and it is deterministic if $Alg$ is deterministic. 

\end{lemma}

\begin{proof}

Fix some integer $k = O(\log (n))$. $Alg'$ will be running $k$ instances of $Alg$ in parallel on graph $G$, call them $A_i : i \in \{0,..,k-1\}$. Assume that $Alg$'s running time is $k$ batch-dynamic. We will describe what state each instance of $Alg$ will take during processing specific parts of the input, then argue that at least one of them will be outputting an $\alpha$ (or $(\alpha, \delta)$)-approximate matching at all times.

Assume that the input sequence $I$ is $k^{k}$ long. Let $I[i]$ represent the $i$-th element of the input sequence and $I[i,j)$ represents elements $i, i+1,...,j-1$ for $j>i$. Let $I[i,j]$ represent $I[i,j) \cup I[j]$. Fix a specific instance of $Alg$ say $A_i$. Call the input batches of $A_i$ as $B^j_i : j \in [k]$. At a given point in time let $|B^j_i|$ refer to the number of input elements instance $A_i$ has progressed as it's $j$-th batch. Note that we will assume that in update time $O(T(n) \cdot |B^j_i|)$ instance $A_i$ can revert back to a state where input batch $B^j_i$ was empty given the elements of $B^j_i$ where the last elements of $I$ progressed by $A_i$. 

Represent the input elements of $I$ as $k$-long $k$-airy strings starting from $\{0\}^{k}$. Choose $I[\lambda]$ such that $\lambda$-s $k$-airy representation ends with an $'i'$ followed by $\gamma > 0$ $'0'$-s and contains a single $i$ digit. We will now describe what instance $A_i$ will be doing while $Alg'$ is processing input elements $I[\lambda,\lambda + k^\gamma)$. We will call this process as the \textit{resetting of batches $B_i^{\lambda},..,B_i^1$}. 

\medskip

\textbf{Resetting the Contents of Batches $B_i^{\lambda},..,B_i^1$}:

\medskip

With a slight overload of notation partition the input sub-sequence $I[\lambda, \lambda + k^{\gamma})$ into $\gamma + 1$ sub-sequences $I_j : j \in \{0,...,\gamma\}$. Let $\lambda_j = \lambda + \sum_{x = j}^{\gamma - 1} k^{j} \cdot (k-1)$ for $\gamma \geq j \geq 0$. Let $I_j = I[\lambda_j, \lambda_{j-1})$ for $\gamma \geq j > 0$ and $I_0 = I[\lambda_0]$. Observe that $|I_j| = \Theta(k^j)$.

\begin{itemize}
    \item While $Alg'$ is processing input elements $I_\lambda$ instance $A_i$ will revert to the state it was in before processing the contents of the batches $B_i^{\gamma + 1},...,B_i^1$. Then it proceeds to place all these elements into batch $B_i^{\gamma + 1}$ a single batch.
    
    \item While $Alg'$ is processing input elements $I_j : \gamma > j > 0$ instance $A_i$ will progress input elements $I_{j+1}$ as batch $B_i^{j+1}$. 
    
    \item While $Alg'$ is processing the input element $I_0$ instance $A_i$ will place input elements $I_1 \cup I_0$ into $B_i^1$.
\end{itemize}

If $A_i$ is not resetting batches it is processing single elements of the input string.

\medskip

\textbf{Processing Single Elements of The Input String:}

\medskip

If the first $k-1$ digits of the $k$-airy representation of $\lambda$ don't contain a single $i$ digit then while $Alg$ is processing $I[\lambda]$ instance $A_i$ will extend it's last batch $B_i^1$ with input element $I[\lambda]$.

\medskip

These two instances describe the behaviour of $A_i$ over the whole of $I$. If $A_i$ is processing a single input element at any point in time it's output is an $\alpha$ (or $(\alpha, \delta)$)-approximate matching. Also observe, that for any $\lambda$ there is a digit $i \in [k]$ in it's $k$-airy representation which is not one of it's first $k-1$ digits. By definition, this implies that $A_i$ will be be processing $I[\lambda]$ as a single input element. Hence, the output of $A_i$ will be an $(\alpha)$ (or $(\alpha, \delta)$)-approximate matching for some $i$ at all time steps.

\begin{claim}

\label{cl:batchsize}

At all times for all $j \in [k]$ and $i \in \{0,...,k-1\}$ it holds that $|B_i^j| \leq (j+1) \cdot k^j$.

\end{claim}

\begin{proof}

We will proof the claim through induction on $j$. Fix $i$. Whenever the contents of $B_i^1$ are reset it will be set to contain exactly $k$ input elements. If the contents of $B_i^1$ are not reset while $I[\lambda]$ is progressed by $Alg'$ then $B_i^1$ is extended by $I[\lambda]$. However, over the course of $k$ consecutive input elements being progressed by $Alg'$ batch $B_i^1$ must be reset. Therefore, $B_i^1$ will never contain more than $2 \cdot k - 1$ elements.

Assume that $|B_i^j| \leq (j+1) \cdot (k^j - k^{j-1})$ at all times as an inductive hypothesis. Consider how many elements may $B_i^{j+1}$ contain. Whenever $B_i^{j+1}$ is reset it will be set to contain exactly $(k-1) \cdot k^{j}$ elements. Furthermore, whenever $B_i^j$ is reset $B_i^{j+1}$ is extended by the contents of $B_i^j,..,B_i^1$. These are the only cases when $B_i^{j+1}$ may be extend by any input elements. $B_i^j$ is reset at most $k-1$ times between two resets of $B_i^{j+1}$. Therefore, at all times $|B_i^{j+1}| \leq (k-1) \cdot (k^j + \sum_{x = 1}^j (x+1) \cdot (k^x - k^{x-1})) \leq (k - 1) \cdot (j + 2) \cdot k^{j} = (j + 2) \cdot (k^{j+1} - k^j)$. This finishes the inductive argument.

\end{proof}

\begin{claim}

The worst-case running time of $A_i$ is $O(T(n) \cdot k^2)$ for all $i \in \{0,...,k-1\}$.

\end{claim}

\begin{proof}

To bound worst case running times differentiate two cases. Firstly, if $I[\lambda]$ is progressed as a single input element by $A_i$ then $A_i$ will extend it's smallest batch $B_i^1$ with $I[\lambda]$. As at all times $|B_i^1| \leq 2 \cdot k$ due to Claim~\ref{cl:batchsize} this can be done in worst-case update time $O(T(n) \cdot k)$.

Fix $\lambda$ as described previously, such that it's $k$-airy representation contains a single $i$ digit followed by $\gamma > 0$ 0-s so that $A_i$ will be resetting batches $B_i^{\gamma},..,B_i^1$ while $Alg'$ is processing $I[\lambda, \lambda + k^\gamma)$. Define $I_j : \gamma \geq j \geq 0$ as before. While $Alg'$ is processing $I_\gamma$ instance $A_i$ has to revert to the state before processing any of $B_i^{\gamma + 1}, ..., B_i^1$ and progress their contents as a single batch into $B_i^{\gamma + 1}$. This concerns the backtracking and processing of $O(k^{\gamma+1} \cdot \gamma)$ input elements by Claim~\ref{cl:batchsize}. The computational work required to complete this can be distributed over the time period $Alg'$ is handling $I_\gamma$ evenly as this computation doesn't require $A_i$ to know the contents of $I_\gamma$. Hence, it can be completed in $O(T(n) \cdot k \cdot \gamma) = O(T(n) \cdot k^2)$ worst-case update time.

Similarly, over the course of $Alg'$ processing $I_j$ which consists of $\Theta(k^j)$ elements we can distribute the $O(T(n) \cdot k^{j+1})$ total work of processing $I_{j+1}$ into batch $B_i^{j+1}$ evenly resulting in $O(T(n) \cdot k)$ worst case update time. Finally, for instance $A_i$ processing $I_1 \cup I_0$ while $Alg'$ progresses $I_0$ will take $O(T(n) \cdot k)$ time.

\end{proof}

Therefore, each instance $A_i$ runs in $O(T(n) \cdot k^2)$ worst-case update time. As there are $k$ instances of $Alg$ running like as described in parallel, this takes a total of $O(T(n) \cdot k^2)$ worst case update time. It remains to select $k = O(\log(n))$ so the algorithm can progress an input of length $O(\log^{\log (n)}(n)) = O(n^{\log (\log (n))})$, that is of input sequences of arbitrarily large polynomial length for large enough $n$.

\end{proof}

\section{Missing Proofs from Section~\ref{sec:Vertex Set Sparsification}}

\subsection{Proof of Lemma~\ref{lm:vertexsparsification:general}}

\begin{algorithm}[H]
	\SetKwInput{KwInput}{Input}                
	\SetKwInput{KwOutput}{Output}              
	\DontPrintSemicolon
	\label{alg:vertexsparsification}
	\caption{Vertex Sparsification}
	
	\KwInput{$G = (V,E), Alg_M, Alg_S$}
	\KwOutput{$(\alpha + \epsilon)$-approximate maximum matching of $G$}
	$MMSize \leftarrow 1$ \;
	$i \leftarrow 1$ \;
    \While{$MMSize \leq n$}
    {
        $i \leftarrow i + 1$ \;
        $MMSize \leftarrow MMSize \cdot (1 + \epsilon/(8\alpha))$ \;
        $\V^i : V^i_j : j \in [L] \leftarrow$ Set of $(MMSize,\epsilon/(8\alpha))$ matching preserving vertex partitionings of $G$ into $C\cdot MMSize$ vertex subsets of size L output by $Alg_S$ \;
        $G^i_j \leftarrow$ Vertex concatenation of $G$ based on $V^i_j$ \;
        $M^i_j \leftarrow$ Maintain $(\alpha, \frac{\epsilon}{8C})$-approximate matching of $G^i_j$ with $Alg_M$ \;
        $G_{M^i_j} \leftarrow$ Maintain edges of $M^i_j$ in $G$ \;
    }
    $E' \leftarrow \underset{i,j}{\cup} G_{M^i_j}$ \;
    $M^* \leftarrow $ Maintain a $1 + \epsilon/(8\alpha)$-approximate maximum matching of $(V,E')$ with Lemma~\ref{lm:guptapeng}\;
    
\end{algorithm}

\begin{claim}

Algorithm~\ref{alg:vertexsparsification} maintains an $(\alpha + \epsilon)$-approximate maximum matching. 

\end{claim}

\begin{proof}

Fix $i = \floor{\log_{1 + \epsilon/(8\alpha)}(\mu(G))}$ and let $\mu_{1 + \epsilon/(8\alpha)}(G) = (1 + \epsilon/(8\alpha))^i$. Note that $G$ contains a matching of size $\mu_{1 + \epsilon/(8\alpha)}(G)$ (assume integrality for sake of convenience) and $\mu(G) \leq \mu_{1 + \epsilon/(8\alpha)}(G) \cdot (1 + \epsilon/(8\alpha))$. By the definition of matching preserving vertex partitionings there is a $j \in [L]$ such that $\mu(G^i_j) \geq (1 - \epsilon/(8\alpha)) \mu_{1 + \epsilon/8}(G)$. 

Hence, $\mu(G^i_j) \geq \mu(G) \cdot (1 - \epsilon/(4\alpha))$. As the vertex set of $G^i_j$ is of size $C \cdot \mu_{1 + \epsilon/8}(G)$ we have that $|M_i^j| \cdot \alpha + \frac{\epsilon}{8C} \cdot C \cdot \mu_{1 + \epsilon/(8/\alpha)}(G) \geq \mu(G_i^j) \geq \mu(G) \cdot (1 - \epsilon/(4\alpha))$ as $M^i_j$ is an $(\alpha, \frac{\epsilon}{8C})$-approximate maximum matching of $G^i_j$. This simplified states that $|M^i_j| \cdot \frac{\alpha}{1 - \frac{3\cdot \epsilon}{\alpha \cdot 8}} \geq \mu(G)$.

As $M^i_j \subseteq E'$ we have $|M^*| \cdot (1 + \epsilon/(8\alpha)) \geq |M^i_j|$ and therefore $|M^*| \cdot \frac{\alpha \cdot (1 + \frac{\epsilon}{8 \alpha})}{1 - \frac{\epsilon \cdot 3}{8 \cdot \alpha}} \geq \mu(G)$. This can be simplified to $|M^*| \cdot (\alpha + \epsilon) \geq \mu(G)$.

\end{proof}

\begin{claim}

Algorithm~\ref{alg:vertexsparsification} has an update time of $O(T(n, \epsilon/C) \cdot \frac{L^2 \cdot \log^2(n)}{\epsilon^4})$.

\end{claim}

\begin{proof}

The maintenance of $M^i_j$ will take $O(T(n, \epsilon/C))$ update time for specific values of $i,j$. As $\alpha = O(1)$ $i$ will range in $[O(\frac{\log(n)}{\epsilon})]$. Therefore, the algorithm maintains $O(\frac{L \cdot \log(n)}{\epsilon})$ matchings in parallel using $Alg_M$. This means $E'$ has maximum degree $O(\frac{L \cdot \log(n)}{\epsilon})$ and can be maintained in update time $O(T(n,\epsilon/C) \cdot \frac{L \cdot \log (n)}{\epsilon})$ and may undergo the same amount of recourse. Hence, with the invocation of the algorithm from Lemma~\ref{lm:guptapeng} the total update time is $O(T(n,\epsilon/C) \cdot \frac{L^2 \cdot \log^2}{\epsilon^4})$.

The two claims conclude Lemma~\ref{lm:vertexsparsification:general}

\end{proof}

Do note, that the update time can be slightly improved to $O(T(n, \frac{\epsilon}{C}) \cdot \frac{L \cdot \log (n)}{\epsilon} + \frac{L^2 \cdot \log^2(n)}{\epsilon^5})$ using Lemma~\ref{lm:recoursebound} (\cite{recoursebound}). The update time of the sparsifier is $O(T(n, \frac{\epsilon}{C}) \cdot \frac{L \cdot \log (n)}{\epsilon})$. Using the lemma it's recourse can be bounded at $O(\frac{L \cdot \log(n)}{\epsilon})$. Applying Lemma~\ref{lm:guptapeng} (\cite{gupta2013fully}) yields the slightly different update time.

\subsection{Proof of Lemma~\ref{lm:partition:randomized}}

\label{app:sparsification}

\begin{proof}

For graph $G = (V,E)$ generate $L = \ceil{\frac{512 \cdot \log (n)}{\epsilon^2}}$ vertex partitionings into $d = \ceil{4 \cdot \frac{(2k)}{\epsilon}}$ sets at random. Call the set of partitionings $\V = \{ \V^j : j \in [L]\}$ and let $V_i^j$ stand for the $i$-th vertex set of the $j$-th partitioning. Fix $2k$ vertices $S$ of $V$ arbitrarily to represent the endpoints of a matching of size $k$ in $G$ and note that this can be done $\binom{n}{2 \cdot k} \leq n^{2 \cdot k} \leq e^{\ln (n) \cdot 2 \cdot k} \leq e^{4 \cdot \log_2 (n) \cdot k}$ number of ways.

Fix a specific vertex partitioning $\V^j$ with vertex sets $V_i^j : i \in [d]$. Let the random variable $X^j_i : i \in [d]$ be an indicator variable of $S \cap V_i^j \neq \emptyset$ and $\bar{X^j} = \sum_{i \in [d]} X^j_i$.

\begin{claim}

$X_i^j : i \in [d]$ are negatively associated random variables.

\end{claim}

\begin{proof}

Define $B_i^l : i \in [d], l \in [2 \cdot k]$ be the indicator variable of the $l$-th vertex of $S$ falling into the $i$-th subset $V^j_i$. This turns the random variables into the well known balls and binds experiment. By \cite{ballsbins1} (this can also be considered a folklore fact) random variables $B_i^l : i \in [d], l \in [2 \cdot k]$ are negatively associated. By definition $X_i^j = \max_{l \in [2 \cdot k]} \{ B_i^l\}$. By Theorem 2 of \cite{ballsbins2} monotonously increasing functions defined on disjoint subsets of a set of negatively associated
random variables are negatively associated. As $\max$ is monotonously increasing this implies that $X_i^j : i \in [d]$ are also negatively associated.

\end{proof}

$$\mathbb{E}[X^j_i] = 1 - \Pr[S \cap V_i^j = \emptyset] = 1 - (1 - \frac{1}{d})^{2k} \geq 1 - ((1 - \frac{1}{8 \cdot k /\epsilon})^{8 \cdot k/\epsilon})^{\epsilon/4} \geq 1 - e^{-\epsilon/4} \geq \frac{\epsilon \cdot (1 - \epsilon/8)}{4}$$

Therefore, $\mathbb{E}[\bar{X^j}] \geq \ceil{\frac{8 \cdot k}{\epsilon}} \cdot \frac{\epsilon \cdot (1 - \epsilon/8)}{4} \geq 2k \cdot (1 - \epsilon/8)$. Now we apply Chernoff's inequality for negatively associated random variables to get that:

$$\Pr[\bar{X^j} \leq 2k \cdot (1 - \epsilon/4)] \leq \Pr[\bar{X^j} \leq \mathbb{E}[\bar{X^j}] \cdot (1 - \epsilon/8)] \leq \exp (-\frac{\mathbb{E}[\bar{X^j}] \cdot (\frac{\epsilon}{8})^2}{2}) \leq e^{\frac{-2k \cdot \epsilon^2}{128}}$$

This implies that 

$$\Pr[\min_{j \in [L]} \{ \bar{X}^j \} \leq 2k \cdot (1 - \epsilon/4)] \leq e^{-4 \cdot \log(n) \cdot 2k}$$

Further applying a union bound over the $\binom{n}{2k}$ possible choices of $S$ yields that regardless of the choice of $S$ with probability $1 - e^{ - 2 \cdot \log (n) \cdot 2k} \geq 1 - 1/poly(n)$ there is a partitioning $\V^j = V^j_i : i \in [d]$ where at least $2k \cdot (1 - \epsilon/4)$ of the vertex sets of $\V^j$ contain a vertex of $S$. This implies that there can be at most $2k \cdot (1 - \epsilon/2)$ vertices of $S$ sharing a vertex set of $\V^j$ with an other vertex of $S$. Furthermore, if $S$ represents the endpoints of a matching of size $k$ at least $k \cdot (1 - \epsilon)$ of it's edges will have both their endpoints being assigned to unique vertex sets of $\V^j$ with respect to $S$. This implies that the concatenation of $G$ based on $\V^j$ will preserve a $1 - \epsilon$ fraction of any matching of size $k$ from $G$. Therefore, $\V$ is a $(k,\epsilon)$ matching preserving set of partitionings for $G$.

Note that while we can simply sample the partitionings randomly in polynomial time, we could also consider all possible sets of partitionings and check weather any of them is $(k, \epsilon)$ matching preserving for all possible choice of $S \subseteq V$ . From the fact that a random sampling based approach succeeds with positive probability we know that there is a set of $(k, \epsilon)$ matching preserving partitionings therefore we will find one one eventually deterministically.

\end{proof}

\section{Missing Proofs of Section~\ref{sec:damagedEDCS}}

\label{app:EDCS}

\subsection{Proof of Lemma~\ref{lm:static:damagedEDCS:correctness}}

\begin{proof}

Let $E_{fin}'$ represent the state of $E'$ at termination. First let's argue that $\forall e \in H_{fin} : deg_{H_{fin}} \leq \beta$. At the end of the penultimate iteration of the outer loop $H$ must have maximum edge degree of $\beta \cdot (1 - \lambda/4)$. $H$ then will be extended with edges of $E' \setminus E_D$ which has a maximum degree of $\beta \cdot \lambda / 8$. Therefore, $\max_{e \in H_{fin}} deg_{H_{fin}}(e) \leq \beta \cdot (1 - \lambda/4) + 2 \cdot \lambda/8 \leq \beta$.

As $\sum_{v \in V}deg_{E_{fin}'}(v) \leq \frac{\delta \cdot \lambda \cdot \beta \cdot n}{8}$ it must hold that $|V_D| \leq \delta \cdot n = |V| \cdot \delta$. Take an edge $e \in E \setminus H_{fin}$ which doesn't intersect $V_D$. As all such edges with lower than $\beta \cdot (1 - \lambda/2)$ edge degree in $E$ were added to $E_{fin}'$ it must hold that $deg_{H_{fin} \cup E_{fin}'}(e) \geq \beta \cdot (1 - \lambda/2)$. As neither endpoints of $e$ are in $V_D$ it must hold that $deg_{E_D}(e) \leq \lambda \cdot \beta / 4$. This implies that $deg_{H_{fin}}(e) \geq deg_{H_{fin} \cup E_{fin}'}(e) - deg_{E_D}(e) \geq \beta \cdot (1 - \lambda/2) - \lambda \cdot \beta /4 \geq \beta \cdot (1 - \lambda)$. Hence, $H_{fin}$ is a $(\beta, \lambda, \delta)$-damaged EDCS of $G$.

\end{proof}

\subsection{Proof of Lemma~\ref{lm:static:damagedEDCS:runningtime}}

\begin{proof}

Observe that every iteration of the repeat loop runs in $O(m)$ time as each iteration can be executed over a constant number of passes over the edge set. Define $\Phi(H) = \Phi_1(H) - \Phi_2(H)$ where $\Phi_1(H) = \sum_{v \in V} deg_H(v) \cdot (\beta - 1/2) = |E(H)| \cdot (2\cdot \beta -1)$ and $\Phi_2(H) = \sum_{e \in H} deg_H(e)$. Initially $\Phi(H) = 0$ and $\Phi(H) \leq \beta^2 \cdot n$. We will show that $\phi(H)$ monotonously increases over the run of the algorithm and each iteration of the repeat loop (except for the last one) increases it by at least $\Omega (\beta^2 \cdot \lambda^2 \cdot \delta \cdot n)$ which implies the lemma.

$\Phi(H)$ may change at times when edges are added to or removed from $H$. Whenever $e$ is removed from $H$ we know that $deg_H(e) > \beta \cdot (1 - \lambda/4)$ (before the deletion). This means that $\Phi_1(H)$ decreases by $2\beta \cdot (1 - \lambda/4) - 1$ but $\Phi_2(H)$ also decreases by at least $2 \cdot \beta \cdot (1 - \lambda/4)$. This is because $deg_H(e)$ disappears from the sum of $\Phi_2(H)$ and $deg_H(e) - 2$ elements of the sum (degrees of edges neighbouring $e$) reduce by $1$ and $deg_H(e) \geq \beta \cdot (1 - \lambda/4) + 1$. Hence, $\Phi(H)$ increases by at least 1.

Whenever an edge $e$ is added to $H$ we know that $deg_H(e) < \beta \cdot (1 - \lambda/2)$ (before the insertion). Due to the insertion $\Phi_1(H)$ increases by exactly $2 \cdot \beta - 1$. $\Phi_2(H)$ increases by at most $2 \cdot \beta \cdot (1 - \lambda/2)$ as a term of at most $\beta \cdot (1 - \lambda/2) + 1$ is added to it's sum and at most $\beta \cdot (1 - \lambda/2) - 1$ elements of it's sum increase by $1$. Therefore, $\Phi(H)$ increases by at least $\lambda \cdot \beta$. In every iteration but the last one of the repeat loop at least $\frac{\lambda \cdot \beta \cdot \delta \cdot n}{16}$ edges were added to $H$. This means every iteration increases $\Phi(H)$ by at least $\frac{\lambda^2 \cdot \beta^2 \cdot \delta \cdot n}{16}= \Omega(\lambda^2 \cdot \beta^2 \cdot \delta \cdot n)$ finishing the lemma.

\end{proof}

\subsection{Proof of Lemma~\ref{lm:dynamic:approxEDCS:correctness}}

\begin{proof}

Every time $H$ is reset rebuilt through StaticDamagedEDCS the lemma statement is satisfied (by Lemma~\ref{lm:static:damagedEDCS:correctness}) Focus on one period of $\alpha$ updates after a rebuild. Define $E_D$ and $E_I$ to be the set of edges deleted and inserted over these updates respectively (note that $E_D \cap E_I$ might not be empty). Define $V_{E_D} = \{v \in V|deg_{E_D}(v) \geq \frac{\beta \cdot \lambda}{16}\}$ and $V_{E_I} = \{v \in V|deg_{E_I}(v) \geq \frac{\beta \cdot \lambda}{16}\}$. Note, that $|V_{E_D} \cup V_{E_I}| \leq \frac{2\cdot \alpha}{\frac{\beta \cdot \lambda}{16}} \leq \frac{\delta \cdot n}{2}$.

As after a call to Algorithm~\ref{alg:static:approxEDCS} the sparsifier $H$ is a $(\frac{\beta}{1 + \lambda/4}, \lambda/4,\delta/2)$-damaged EDCS (follows from Lemma~\ref{lm:static:damagedEDCS:correctness}) and  the following holds for some $V_D \subseteq V$ with $|V_D| \leq |V| \cdot \delta/2$:

\begin{itemize}
    \item $\forall e \in H: deg_H(e) \leq \frac{\beta}{1 + \lambda/4}$
    \item All $e \in E \setminus H$ such that $e \cap V_D = \emptyset$ satsifies that $deg_H(e) \geq \frac{\beta \cdot (1 - \lambda/4)}{1 + \lambda/4}$
    
\end{itemize}

Define $V_D' = V_D \cup V_{E_D} \cup V_{E_I}$. Note that $|V_D'| \leq |V_D| + |V_{E_D} \cup V_{E_I}| \leq n \cdot \delta$. Also note that after a rebuild $\max_{e \in E} deg_H(e) \leq \frac{\beta}{1 + \lambda/4}$. As edges will only be inserted between vertices $u$ and $v$ if their degrees is at most $\frac{\beta \cdot \lambda}{16} - 2$ in $H$ we can be certain that at any point $\max_{e \in E} deg_H(e) \leq \frac{\beta}{1 + \lambda/4} +  \frac{\beta \cdot \lambda}{16} \leq \beta$ (for small enough values of $\lambda$).

At any point during the phase take an arbitrary $e \in E \setminus H \land e \cap V_D' = \emptyset$. If $e \in E_I$ at the time of it's (last) insertion one of its endpoints, say $v$ had $deg_{E_I}(v) \geq \frac{\lambda \cdot \beta}{16}$ or $deg_H(e) > \beta -2$. The former would imply $v \in V_D'$. Therefore, we can assume that if $e \notin H$ then either $e \in E_I$ and at time of it's insertion $deg_H(e) > \beta - 2$ or $e \notin E_I$ and at the start of the phase $deg_H(e) \geq \beta \frac{ (1 - \lambda/4)}{1 + \lambda/4}$. Either way, during the phase the edge degree of $e$ may have reduced by at most $\frac{\beta \cdot \lambda}{8}$ as none of it's endpoints are in $V_{E_D}$. Therefore, $deg_H(E) \geq \frac{\beta \cdot (1 - \lambda/4)}{1 + \lambda/4} - \frac{\beta \cdot \lambda}{8} \geq \beta \cdot (1 - \lambda)$. This concludes the proof.

\end{proof}

\subsection{Proof of Lemma~\ref{lm:dynamic:damagedEDCS:amortizedbounds}}

\begin{proof}

Edge insertions and deletions are handled in $O(1)$ time apart from the periodic rebuilds. The rebuilds run in $O(\frac{m}{\delta \cdot \lambda^2})$ deterministic time by Lemma~\ref{lm:static:damagedEDCS:runningtime} therefore over $\alpha$ insertions the amortized update time is $O(\frac{m}{\delta \cdot \lambda^2 \cdot \alpha}) = O(\frac{m}{n \cdot \beta \cdot \lambda^3 \cdot \delta^2})$. The recourse of the sparsifier is also constant apart from rebuild operations. When a rebuild occurs the sparsifier goes under at most $O(n \cdot \beta)$ edge updates. Therefore, the amortized recourse is $O(\frac{n \cdot \beta}{\alpha}) = O(\frac{1}{\lambda \cdot \delta})$.

\end{proof}

\section{Proof of Lemma~\ref{lm:BKbatch}}

\label{app:BK-kbatch}

In order to prove Lemma~\ref{lm:BKbatch} we will need to make small modifications to Algorithm 4 and Algorithm 6 of \cite{bhattacharya2021deterministic}. For the convenience of the reader exact copies of Algorithm 1-6 from \cite{bhattacharya2021deterministic} are added here (as Algorithm~\ref{alg:uniform:sparsify} to Algorithm~\ref{alg:deletion}). The complete algorithm of \cite{bhattacharya2021deterministic} is quite long therefore here we will just be focusing on the changes we have to make in order to turn it $k$ batch-dynamic. In order to see how these changes integrate into the complete framework of \cite{bhattacharya2021deterministic} we will explain their algorithm on a high-level.

A fractional matching $w$ is a function mapping the edges of a graph to the $[0,1]$ interval, $f:e \rightarrow [0,1]$, such that the sum of the weights of the edges incident on any vertex adds up to at most one, $\forall v \in V \sum_{e \in N_E(v)} w(e) \leq 1$. The size of a fractional matching is just the sum of the weights on all its edges: $size(w) = \sum_{e \in E} w(e)$. There is a close connection between the maximum fractional and integral matching size of graph $G$. It's folklore knowledge that if $G$ is bipartite then the maximum size of integral and fractional matchings on $G$ are the same. If $G$ is not bipartite then the maximum fractional matching size is at most $3/2$ as large as the maximum integral matching. Therefore, if an algorithm can maintain a sparse $(\alpha)$-approximate fractional matching sparsifier $H$ in the case of bipartite graphs $H$ is also an $(\alpha)$-approximate integral matching sparsifier (for general graphs some more technical effort is required as shown in \cite{bhattacharya2021deterministic}).

The algorithm of \cite{bhattacharya2021deterministic} for bipartite graphs can be (on a very high level) described as follows: 

Step 1 : Maintain a $(2 + \epsilon)$-approximate fractional matching $w$ on graph $G$ (note that to handle the case of general graphs additional properties of $w$ are required) using folklore dynamic fractional matching algorithms.

Step 2 : Maintain $E' \subseteq E$ and $w' : E' \rightarrow [0,1]$ such that $E'$ is sufficiently sparse while $size(w') \geq (1 - O(\epsilon)) \cdot size(w)$.

Step 3 : Maintain a $(1+\epsilon)$-approximate integral maximum matching in $E'$.

\medskip

In this process Step 1 is executed using the algorithm of \cite{bhattacharya2019deterministically} while Step 3 is using the algorithm of \cite{peleg2016dynamic}. Both of these algorithms have worst-case update time bounds therefore they are also $k$ batch-dynamic for any $k$. Step 2 of this process involves amortization which we will relax into $k$ batch-dynamic.

Step 2 is executed in two phases. Firstly, in phase A) $w$ is discretized. This achieved through defining fractional matching $w_r$ such that $w_r(e)$ is the smallest power of $(1 + \epsilon)^{-1}$ smaller then $w(e)$ (edges with very small, less than $\epsilon/n^2$ weight are ignored). By definition $w_r$ is the union of $O(\frac{\log(n)}{\epsilon})$ uniform fractional matchings and $size(w_r) \cdot (1 + \epsilon) \geq size(w)$. The uniform fractional matchings are then are maintained separately and their union is returned as the sparsifier. 

In phase B) the algorithm sparsifies the $O(\frac{\log(n)}{\epsilon})$ uniform fractional matchings in parallel. Specifically, assume $w_\lambda : E_{\lambda} \rightarrow \lambda$ is one of these uniform fractional matchings. Taking $w_\lambda$ as an input Algorithms~\ref{alg:uniform:sparsify} to \ref{alg:deletion} maintain $w'_\lambda : E_{\lambda}' \rightarrow [0,1]$ such that: 

\begin{itemize}

    \item $\forall e \in E'_{\lambda} : w_{\lambda}'(e) < \beta$
    \item $\forall v \in V : w_{\lambda}'(v) \leq w_{\lambda}(v)$
    \item $size(w_\lambda) \leq (1 + O(\epsilon \cdot \log(\beta/\lambda))) \cdot size(w_{\lambda}')$ 

\end{itemize}
for constants $\lambda, \beta$ (by Lemma 4.4 of \cite{bhattacharya2021deterministic}).Furthermore, the arboricity of $E_{\lambda}'$ is $O(\frac{1}{\epsilon} + \frac{1}{\beta})$ and $E_{\lambda}'$ is maintained in $O(\epsilon \cdot \log(\frac{\beta}{\lambda}))$ amortized update time (Lemma 4.3 and Lemma 4.5 of \cite{bhattacharya2021deterministic}).

As phase A), the discretization of $w$, can be done in $O(1)$ worst-case update time we will be focusing on the $k$ batch-dynamization of phase B). We will show that through simply adjusting the slack parameters we will be able to guarantee all properties of $E_{\lambda}'$ and $w_{\lambda}'$ above while changing the running time from amortized $O(\epsilon \cdot \log(\frac{\beta}{\lambda}))$ to $k$ batch amortized $O(k \cdot \epsilon \cdot \log(\frac{\beta}{\lambda}))$. Substituting the modified version of phase B) into the framework of \cite{bhattacharya2021deterministic} we therefore get a $k$ batch algorithm with an update time which is an $O(k)$-factor slower.

The only modifications we need to make concern the lines highlighted in red from Algorithm~\ref{alg:insertion} and Algorithm~\ref{alg:deletion}. Partition input sequence $I$ into batches $I_i : i \in [k]$. Administer the following modification (highlighted with blue text): while progressing batch $i$ exchange slack parameter $\epsilon$ with $\epsilon_i = \frac{\epsilon \cdot i}{k}$.

As the modified algorithm will be operating with slack parameters strictly tighter at all times all requirements of Lemmas 4.3 and 4.4 of \cite{bhattacharya2021deterministic} will be enforced at all times. The only challenge is to argue about the update time of the algorithm.

Firstly, let's consider how much time will handling edge insertions through Algorithm~\ref{alg:insertion} will take. Assume that the progressing of input batch $I_{i-1}$ has finished at time $\tau_0$ and the algorithm currently halts before starting progressing the elements of input batch $I_i$. By the description of Algorithm~\ref{alg:insertion}, at this point in time $|E_{(p)}| \leq \frac{\epsilon \cdot (i-1)}{k} \cdot |E_{(a)}|$ as enforced while progressing batch $I_{i-1}$. The next time Algorithm~\ref{alg:insertion} is triggered to do more than $O(1)$ work is when $|E_{(p)}|$ exceeds $\frac{\epsilon \cdot i}{k} \cdot |E_{(a)}|$, call this a critical event. As during a single edge insertion or deletion either $|E_{(p)}|$ increases by one (in case of an insertion) or $|E_{(a)}|$ decreases by one (in case of a deletion) there will be at least $O(|E_{(a)}|) \cdot \frac{\epsilon}{k}$ edge updates before a critical event. 

Hence, if such an event doesn't occur during the progressing of batch $I_i$ then the handling of insertions will require $O(|I_i|)$ time over the whole batch. If such an event occurs at time $\tau$ then a call to Static-Uniform-Sparsify and Clean-Up will be made. Both of these subroutines will run in $O(|E_{(a)}|)$ total time (by Lemma 3.3 of \cite{bhattacharya2021deterministic} and by observation). Therefore, to handle updates from time $\tau_0$ to $\tau$ the algorithm will require $O(\tau - \tau_0 + |E_{(a)}|) = O((\tau - \tau_0) \cdot \frac{k}{\epsilon})$ time. After the update at $\tau$ is progressed the invariant $|E_{(a)}| \cdot \frac{\epsilon \cdot (i-1)}{k} \geq |E_{(p)}|$ is again satisfied (as $E_{(p)}$ is emptied by Clean-Up) therefore we can inductively argue that the processing of the whole batch will take $O(|I_i| \cdot \frac{k}{\epsilon})$ time. 

Similarly, let's consider how much update time handling deletions through Algorithm~\ref{alg:deletion} takes. Again assume that at time $\tau_0$ batch $i-1$ has just been progressed. Note that $L = O(\log(\frac{\beta}{\lambda}))$. Firstly, given the deletion of edge $e$ we might need to add it to edge sets $D^{(\geq i)}$ which can take up to $O(L)$ time. The time required to handle deletions will be dominated by the time it will take to run Clean-Up and Rebuild subroutines given the If statement (highlighted with blue) is satisfied. The if statement considers $L+1$ different layers. Focus on a specific layer $j$. At time $\tau$ we are guaranteed that $|D^{(\geq j)}| \leq \frac{\epsilon \cdot (i-1)}{k} \cdot |E^{(\geq j})|$. An edge deletion may increase $|D^{(\geq j)}|$ by one or reset $|D^{(\geq j)}|$ to $0$ through a rebuild on an other layer. This means that there will be at least $O(\frac{\epsilon}{k} \cdot |E^{(\geq j)}|)$ edge changes before the if statement of level $j$ may be satisfied. At that point Clean-Up and Rebuild will take $O(|E^{(\geq j)}|)$ total time to complete (by Lemma 3.3 of \cite{bhattacharya2021deterministic}) and $|D^{(\geq j)}| \leq \frac{\epsilon \cdot (i-1)}{k} \cdot |E^{(\geq j})|$ is again satisfied as $|D^{(\geq j)}|$ is emptied. Hence, we can again argue that it will take at most $O(|I_i| \cdot \frac{k}{\epsilon})$ total time to satisfies the If condition on the $j$-th layer during input batch $I_i$. Therefore, the total update time of the algorithm over batch $I_i$ is in $O(L \cdot \frac{k}{\epsilon} \cdot |I_i|)$ as required. This finishes the proof.

\subsection{Sub-Routines From \cite{bhattacharya2021deterministic}}

These sub-routines are coppies from \cite{bhattacharya2021deterministic} and are included just for the readers convenience. The changes made are to the rows highlighted with red in Algorithm~\ref{alg:insertion} and Algorithm~\ref{alg:deletion}.

\begin{algorithm}
	\SetKwInput{KwInput}{Input}                
	\SetKwInput{KwOutput}{Output}              
	\DontPrintSemicolon
	\caption{Static-Uniform-Sparsify}
	\label{alg:uniform:sparsify}
	\KwInput{$(G = (V, E),  \lambda)$, where $\delta / n^2 \leq \lambda < \beta$}
	Let $w : E \rightarrow [0, 1]$ be a $\lambda$-uniform fractional matching in $G$ \;
	Initialize $V^{(\geq 0)} := V$ and $E^{(\geq 0)} := E$ \;
	Initialize a weight-function $h^{(0)} : E^{(\geq 0)} \rightarrow [0, 1]$ so that $h^{(0)}(e) := \lambda$ for all $e \in E^{(\geq 0)}$ \;
    Let $L := L(\lambda)$ be the unique nonnegative integer $k$ such that $\beta/2 \leq 2^k \lambda < \beta$ \;
    Call the subroutine Rebuild$(0, \lambda)$ \;
    Define $F := \bigcup_{i=0}^{L} F^{(i)}$, and $H := (V, F)$ \;
    Define  $h : F \rightarrow [0, 1]$ such that for all $i \in [0, L]$ and $e \in F^{(i)}$ we have $h(e) := h^{(i)}(e)$ \;

\end{algorithm}
 
\begin{algorithm}
	\SetKwInput{KwInput}{Input}                
	\SetKwInput{KwOutput}{Output}              
	\DontPrintSemicolon
	\caption{Rebuild}
	\label{alg:rebuild}
	\KwInput{$(i',    \lambda)$}
	\For{$i=i'$ to $\left(L-1\right)$}
	{
		$V^{(i)} \leftarrow \emptyset$ \;
    	\While{there is some node $v \in V^{(\geq i)}\setminus V^{(i)}$ with $\text{deg}_{E^{(\geq i)}}\left(v, V^{(\geq i)} \setminus V^{(i)} \right) \leq (1/\epsilon)$}
    	{
    		$V^{(i)} \leftarrow V^{(i)} \cup \{v\}$ \;
    	}
    	$V^{(\geq i+1)} \leftarrow V^{(\geq i)} \setminus V^{(i)}$ \;
    	$F^{(i)} \leftarrow \{ (u, v) \in E^{(\geq i)} :  \text{ either } u \in V^{(i)} \text{ or } v \in V^{(i)} \}$ \;
    	$E^{(\geq i+1)} \leftarrow \text{{\sc Degree-Split}}(E^{(\geq i)} \setminus F^{(i)})$ \;
        \For{all edges $e \in E^{(\geq i+1)}$}
        {
            $h^{(i+1)}(e) \leftarrow 2 \cdot h^{(i)}(e)$ \;
        }
    }
    $F^{(L)} \leftarrow E^{(\geq L)}$ \;
    $V^{(L)} \leftarrow V^{(\geq L)}$ \;
\end{algorithm}
 
\begin{algorithm}
	\SetKwInput{KwInput}{Input}                
	\SetKwInput{KwOutput}{Output}              
	\DontPrintSemicolon
	\caption{Degree-Split}
	\KwInput{$E'$}
	\label{alg:degreesplit}
	Initialize $E^* \leftarrow E'$ and $\mathcal{W} \leftarrow \emptyset$ \;
	\While{$E^* \neq \emptyset$}
	{
	    Let $G^* := (V(E^*), E^*)$, where $V(E^*)$ is the set of endpoints of the edges in $E^*$ \;
	    Compute a {\em maximal} walk $W$ in $G^*$ \;
	    Set $\mathcal{W} \leftarrow \mathcal{W} \cup \{ W \}$, and $E^* \leftarrow E^* \setminus W$ \;
	}
    Return the set of edges $E'' := \bigcup_{W \in \mathcal{W}} W^{(even)}$ \;
\end{algorithm}

\begin{algorithm}
	\SetKwInput{KwInput}{Input}                
	\SetKwInput{KwOutput}{Output}              
	\DontPrintSemicolon
	\caption{Handle-Insertion}
	\label{alg:insertion}
	\KwInput{$(e, \lambda)$}
	$E_{(p)} \leftarrow E_{(p)} \cup \{e \}$ \;
	\If{\textcolor{red}{$\left| E_{(p)} \right| > \epsilon \cdot \left| E_{(a)} \right|$}} {}
	\If{\textcolor{blue}{$\left| E_{(p)} \right| > \frac{\epsilon \cdot i}{k} \cdot \left| E_{(a)} \right|$ (while progressing input batch $I_i$)}}
	{
        Call the subroutine {\sc Clean-Up}$(0, \lambda)$ \;
        $E_{(a)} \leftarrow E_{(a)} \cup E_{(p)}$ \;
        $E_{(p)} \leftarrow \emptyset$ \;
        Call the subroutine {\sc Static-Uniform-Sparsify}$\left(G_{(a)} := (V, E_{(a)}), \lambda \right)$ \;
	}
\end{algorithm}

\begin{algorithm}
	\SetKwInput{KwInput}{Input}                
	\SetKwInput{KwOutput}{Output}              
	\DontPrintSemicolon
	\caption{Clean-Up}
	\label{alg:cleanup}
	\KwInput{$(j, \lambda)$}
	\For{all $i = j$ to $L$}
	{
        $E^{(\geq i)} \leftarrow E^{(\geq i)} \setminus D^{(\geq i)}$ \;
        $F^{(i)} \leftarrow F^{(i)} \setminus D^{(\geq i)}$ \;
        $D^{(\geq i)} \leftarrow \emptyset$ \;
	}
\end{algorithm}

\begin{algorithm}
	\SetKwInput{KwInput}{Input}                
	\SetKwInput{KwOutput}{Output}              
	\DontPrintSemicolon
	\caption{Handle-Deletion}
	\label{alg:deletion}
	\KwInput{$(e, \lambda)$}
	\If{$e \in E_{(p)}$}
	{
        $E_{(p)} \leftarrow E_{(p)} \setminus \{ e \}$ \;
	}
	\Else
	{
	    $k \leftarrow \ell(e) :=  \max \left\{i \in [0, L]  : e \in E^{(\geq i)} \right\}$ \;
	    $E_{(a)} \leftarrow E_{(a)} \setminus \{e \}$ \;
	    \For{$i = 0$ to $k$}
	    {
	        $D^{(\geq i)} \leftarrow D^{(\geq i)} \cup \{e\}$
	    }
	    \If{\textcolor{red}{$\left| D^{(\geq i)} \right| > \epsilon \cdot \left| E^{(\geq i)} \right|$ for some index $i \in \left[0, L \right]$}} {}
	    \If{\textcolor{blue}{$\left| D^{(\geq i)} \right| > \frac{\epsilon \cdot i}{k} \cdot \left| E^{(\geq i)} \right|$ for some index $i \in \left[0, L \right]$ (while progressing input batch $I_i$)}}
	    {
	        Let $j$ be the minimum index $i \in \left[0, L \right]$ for which $\left| D^{(\geq i)} \right|  > \epsilon \cdot \left| E^{(\geq i)} \right|$ \;
	        Call the subroutine {\sc Clean-Up}$(j, \lambda)$ \;
	        Call the subroutine {\sc Rebuild}$\left(j, \lambda \right)$ \;
	    }
	}
\end{algorithm}

\end{document}